\documentclass{99-Styles/IDS-NF} 
\usepackage{times}
\usepackage{bm}         
\usepackage{amsmath,amssymb,bm,slashed} 
\usepackage{amssymb}    
\usepackage{graphicx}   
\usepackage{verbatim}   
\usepackage{color}      
\usepackage{subfigure}  
\usepackage{hyperref}   
\usepackage{enumerate}

\usepackage{fmtcount}   

\usepackage[square,comma,numbers,sort&compress]{natbib}

\usepackage{bibentry}

\usepackage{enumitem}

\usepackage{textcomp}
\usepackage{bibentry}
\nobibliography*

\begin{document}

\makeatletter

\newcommand{\bra}[1]{\ensuremath{\langle #1 |}}   
\newcommand{\ket}[1]{\ensuremath{| #1 \rangle}}   
\newcommand{\bigbra}[1]{\ensuremath{\big\langle #1 \big|}}   
\newcommand{\bigket}[1]{\ensuremath{\big| #1 \big\rangle}}   
\newcommand{\amp}[3]{\ensuremath{\left\langle #1 \,\left|\, #2%
                     \,\right|\, #3 \right\rangle}}  
\newcommand{\sprod}[2]{\ensuremath{\left\langle #1 |%
                     #2 \right\rangle}}  
\newcommand{\ev}[1]{\ensuremath{\left\langle #1 %
                     \right\rangle}} 
\newcommand{\ds}[1]{\ensuremath{\! \frac{d^3#1}{(2\pi)^3 %
                     \sqrt{2 E_\vec{#1}}} \,}} 
\newcommand{\dst}[1]{\ensuremath{\! %
                     \frac{d^4#1}{(2\pi)^4} \,}} 
\newcommand{\tr}{\text{tr}}
\newcommand{\sgn}{\text{sgn}}
\newcommand{\diag}{\text{diag}}
\newcommand{\BR}{\text{BR}}
\newcommand{\NOvA}{\ensuremath{\mbox{NO}\nu\mbox{A}} }

\renewcommand{\vec}[1]{{\mathbf{#1}}}
\renewcommand{\Re}{{\text{Re}}}
\renewcommand{\Im}{{\text{Im}}}
\newcommand{\iso}[2]{{\ensuremath{{}^{#2}}\ensuremath{\rm #1}}}
\newcommand{\eps}{{\ensuremath{\epsilon}}}
\newcommand{\draftnote}[1]{{\bf\color{red} \MakeUppercase{#1}}}
\newcommand{\panm}[1]{{\color{blue} #1}}
\providecommand{\abs}[1]{\lvert#1\rvert}
\providecommand{\norm}[1]{\lVert#1\rVert}
\newcommand{\gsim}      {\mbox{\raisebox{-0.4ex}{$\;\stackrel{>}{\scriptstyle \sim}\;$}}}
\newcommand{\lsim}      {\mbox{\raisebox{-0.4ex}{$\;\stackrel{<}{\scriptstyle \sim}\;$}}}

\newcommand\brabar{\raisebox{-2.0pt}{\scalebox{.2}{\,\,
\textbf{(}}}\raisebox{-3.25pt}{\scalebox{.8}{{\textendash}}}\raisebox{-2.0pt}{\scalebox{.2}{\textbf{)
}}}}

\def\parenbar{\mathpalette\p@renb@r}
\def\p@renb@r#1#2{\vbox{%
  \ifx#1\scriptscriptstyle \dimen@.7em\dimen@ii.2em\else
  \ifx#1\scriptstyle \dimen@.8em\dimen@ii.25em\else
  \dimen@1em\dimen@ii.4em\fi\fi \offinterlineskip
  \ialign{\hfill##\hfill\cr
    \vbox{\hrule width\dimen@ii}\cr
    \noalign{\vskip-.3ex}%
    \hbox to\dimen@{$\mathchar300\hfil\mathchar301$}\cr
    \noalign{\vskip-.3ex}%
    $#1#2$\cr}}}

%
\providecommand{\anmne}{\mbox{$\bar\nu_{\mu} \rightarrow \bar\nu_e$}} 
\providecommand{\nmne}{\mbox{$\nu_{\mu}\rightarrow\nu_e$}} 
\providecommand{\anm}{\mbox{$\bar\nu_\mu$}} 
\providecommand{\nm}{\mbox{$\nu_\mu$}}
\providecommand{\nue}{\mbox{$\nu_e$}} 
\providecommand{\ane}{\mbox{$\bar\nu_e$}} 
\providecommand{\enu}{\mbox{$E_\nu$}}
\providecommand{\piz}{\mbox{$\pi^0 $}}
\providecommand{\pip}{\mbox{$\pi^+$}} 
\providecommand{\pim}{\mbox{$\pi^-$}}

\parindent 10pt
\pagenumbering{roman}
\setcounter{page}{1}
\pagestyle{plain}

\thispagestyle{empty}
\begin{tabular}{p{5cm} p{5cm} p{6cm}}
  \setlength{\baselineskip}{0.25\baselineskip}
    \leftline{\today}                                         &
    \centering{Final}                                         &
    \rightline{ICFA Neutrino Panel 2014(01)}
    \vspace{-0.4cm}\\
    \hline
\end{tabular}

\begin{center}
  {\bf\LARGE 
    Initial report from the ICFA Neutrino Panel
  }
  \vspace{-0.5cm}
\end{center}

\section*{Executive summary}

\vspace{-0.4cm}
The discovery of neutrino oscillations implies that neutrinos have
mass and that the flavour eigenstates mix.
Non-zero neutrino masses require new physics beyond the Standard Model. 
Revealing the physics behind neutrino masses and mixing is among the highest 
priorities for particle physics in the twenty-first century.

Measurements of the parameters that govern neutrino oscillations will
have a profound impact on our understanding of particle physics,
astrophysics and cosmology.
Such a breadth of impact justifies a far-reaching experimental
programme by which the properties of the neutrino that are presently
unknown are determined, the three-neutrino-mixing picture is tested
and the properties of the neutrino are measured with a precision
sufficient to allow the physics that explains these properties to be
understood. 
Accelerator-based neutrino-oscillation experiments are an essential
component of this programme as they are the only way in which
oscillations between all known flavours can be studied precisely.

The accelerator-based neutrino-oscillation programme is international
in both scope and engagement.
The approved programme will improve our knowledge of the mixing angles
and the mass-squared differences and may be able to determine the
neutrino mass hierarchy. 
In the medium term, the Long-baseline Neutrino Experiment (LBNE) and
the Tokai (J-PARC) to Hyper-Kamiokande experiment 
offer
complementary approaches to searching for the violation of the
matter-anti-matter symmetry in neutrino oscillations.
For these experiments to realise their full potential coordinated
programmes of hadro-production and neutrino-nucleus cross-section
measurements are essential.
A design study of an alternative wide-band beam facility is underway:
the Long-baseline Neutrino Observatory (LBNO) study will report by the
end of 2014.
New concepts such as the implementation of a neutrino beam on the
European Spallation Source (ESSnuSB) and the use of muon decay-at-rest
to search for the matter-anti-matter asymmetry (Daedalus) are also
being studied.

The Neutrino Factory, in which electron- and muon-neutrino beams are
produced from the decay of muons confined within a storage ring, has
been shown to offer the ultimate sensitivity and precision.
The staged implementation of the facility has been studied.
The attractive first stage, ``nuSTORM'', has the potential to make
detailed and precise studies of electron- and muon-neutrino-nucleus
scattering and to make exquisitely sensitive searches for sterile
neutrinos.

A small number of measurements do not fit the elegant
three-neutrino-mixing model.
These measurements can be interpreted as evidence for new ``sterile
neutrino'' states.
If confirmed, these measurements will revolutionise the field.
An experimental programme is underway to investigate these anomalies. 
It is important that these anomalies are addressed.
Short-baseline contributions to this programme should have clear
synergy with the long-baseline programme.

This report presents the conclusions drawn by the Panel from three
regional ``Town Meetings'' that took place between November 2013 and
February 2014. 
The Panel recognises that to maximise the discovery potential of
the accelerator-based neutrino-oscillation programme it will be
essential to exploit the infrastructures that exist at CERN, FNAL and
J-PARC and the expertise and resources that reside in laboratories
and institutes around the world. 
Therefore, in its second year, the Panel will consult with laboratory
Directors, funding-agency representatives, the accelerator-based
neutrino-oscillation community and other stakeholders to:
\begin{itemize}
   \item Develop a road-map for the future accelerator-based
    neutrino-oscillation programme that exploits the ambitions
    articulated at CERN, FNAL and J-PARC and includes the programme of
    measurement and test-beam exposure necessary to ensure the
    programme is able to realise its potential;
  \item Develop a proposal for a coordinated ``Neutrino RD''
    programme, the accelerator and detector R\&D programme required to
    underpin the next generation of experiments; and 
  \item To explore the opportunities for the international
    collaboration necessary to realise the Neutrino Factory.
\end{itemize}

\cleardoublepage

\tableofcontents

\cleardoublepage
\pagenumbering{arabic}                   
\setcounter{page}{1}

\section{Manifesto}
\label{Sect:Intro}

\subsection{Why is neutrino physics important?}
\label{SubSect:Why}

The neutrino is the most abundant matter particle in the Universe;
the number of neutrinos exceeds the number of protons, electrons and
neutrons by a factor approaching ten billion.
So, to understand the Universe we must understand the neutrino.
The neutrino is the least understood matter particle; in contrast to
all other fundamental fermions, some of the neutrino's properties are
unknown while the properties that are known are poorly understood.

The discovery of neutrino oscillations, in which the neutrino type or
flavour changes as the neutrino propagates through space and time,
implies that the neutrino has mass and that the neutrino flavours mix.
Theoretical interpretation of this result requires either that:
\begin{itemize}
  \item Neutrinos are their own antiparticle, in which case they are
    an entirely new form of matter; or
  \item Neutrinos and anti-neutrinos are different particles. This case implies
    that lepton number conservation is a fundamental law of Nature.
\end{itemize}
Hence, neutrino oscillations imply that there are new phenomena beyond
those described by the Standard Model.
Consequently, when considering the programme required to understand
the properties of the neutrino, we must assume nothing but aspire to
a programme of sufficient breadth and precision to elucidate the
underlying phenomena.

Progress since the seminal discovery of neutrino oscillations in 1998
has been rapid.  
Since then, the solar-neutrino anomaly has been resolved, mixing among
all three neutrino flavours has been established and the 
magnitudes of the two mass splittings and the three mixings angles
that characterise the strengths of the mixings have been measured.
To complete our understanding we need to know:
\begin{itemize}
  \item Whether mixing among the three neutrino flavours violates the
    matter-antimatter (CP) symmetry.
    Such leptonic CP-invariance violation (CPiV) would be something
    new and might have cosmological consequences;
  \item Why neutrino masses are so tiny, at least a million times
    smaller than any other known matter particle;
  \item What the ordering of the three neutrino mass eigenstates is.
    While there are constraints on the absolute neutrino-mass scale
    our knowledge of the mass spectrum is incomplete;
  \item Why the strength of mixing among the neutrino flavours is so
    much stronger than the mixing among the quarks; 
  \item Whether empirical relationships between neutrino-mixing
    parameters, or between neutrino- and quark-mixing parameters, can
    be established; and 
  \item Whether the few measurements of neutrino oscillations that are
    not readily accommodated within the elegant framework of
    three-neutrino mixing are statistical fluctuations, systematic
    effects or indications that there is even more to discover.
\end{itemize}

Neutrino physics is important. 
Not only has recent progress in neutrino physics been rapid and
exciting, but the remaining questions to be addressed are fundamental.
The discovery potential of the programme required to address these
questions is substantial.

\subsection{What must the neutrino program do?}
\label{SubSect:What}

If we are to understand neutrinos and their role in, and influence
on, the Universe, we must:
\begin{itemize}
  \item Determine the properties of the neutrino that are presently
    unknown;
  \item Determine whether the three-neutrino mixing picture (the
    ``Standard Neutrino Model'', S$\nu$M) is the whole story; and
  \item Measure the properties of the neutrinos with a precision
    sufficient to allow the physics that explains these properties to
    be understood.
\end{itemize}
This ambitious programme requires a comprehensive set of innovative
experiments that have the potential to make discoveries and to provoke
theoretical progress. 
Accelerator-based neutrino oscillation experiments play a critical
role in this program.
They provide the only means by which both neutrino and anti-neutrino
transitions between all of the three known neutrino flavours can
be studied precisely.
In particular, accelerator-based neutrino-oscillation experiments are
essential to:
\begin{itemize}
  \item Complete our knowledge of the pattern of neutrino masses,
    which will constrain ideas about the underlying physics.
  \item Determine the last remaining unmeasured parameter that describes
    the mixing between the three neutrino flavours, the CP-phase
    $\delta$, and determine whether there is observable CP-invariance
    violation in neutrino oscillations; this would constitute the
    discovery of leptonic CP-invariance violation; 
  \item Clarify the origin of the present anomalies in
    accelerator-based neutrino-oscillation measurements---the few
    measurements that do not seem to conform to the S$\nu$M picture
    would, if established, imply the existence of additional neutrino
    states or interactions, and a mysterious new set of physical phenomena;
  \item Measure transitions between all neutrino and anti-neutrino
    flavours to see whether the three-neutrino mixing picture is
    self-consistent and to check that the oscillations depend only on 
    proper time ($L/E$) rather than depending on the baseline ($L$)
    and neutrino energy ($E$) separately.
    A discovery that the S$\nu$M is inadequate would be a major
    breakthrough; and 
  \item  Make precise measurements of all of the mixing parameters to
    search for clues about the underlying physics. 
    This may be the only way to provoke progress in understanding
    neutrinos.
\end{itemize}

This accelerator-based program is international both in
intellectual interest, in engagement and in scope.
It has tremendous potential to make discoveries that would have
profound impact on our understanding of the physics of fundamental
particles and the evolution of the Universe. 
The international program should encompass:
\begin{itemize}
  \item The timely completion of the present generation of
    accelerator-based neutrino-oscillation experiments;
  \item The implementation of an appropriate programme of new
    ``long-baseline'' and ``short-baseline'' neutrino-oscillation
    experiments accompanied by a measurement programme by which
    systematic uncertainties are reduced such that they are
    commensurate with the statistical power of the oscillation
    experiments; and
  \item The accelerator and detector R\&D programmes required to
    deliver the next generation of facilities culminating in the
    Neutrino Factory which is recognised to be the facility that
    offers the ultimate precision.
\end{itemize}

\section{Review of the status of neutrino oscillations}
\label{Sect:Status}

The neutrino mixing matrix \(U\) translates neutrino-mass eigenstates
into flavour eigenstates through the relation 
\((\nu_e,\nu_\mu,\nu_\tau)^T = U(\nu_1,\nu_2,\nu_3)^T\).
The mixing matrix may be parameterised using three mixing angles 
($\theta_{12}, \theta_{23}, \theta_{13}$), one ``Dirac CP phase''
$\delta$, and two Majorana CP phases ($\alpha_{21}, \alpha_{31}$).
In the three-flavour-mixing framework, $U$ may be written
\cite{Beringer:1900zz}:
\begin{eqnarray}
U &=&
\left(
\begin{array}{ccc}
U_{e1}   & U_{e2}   & U_{e3} \\
U_{\mu1} & U_{\mu2} & U_{\mu3} \\
U_{\tau1} & U_{\tau2} & U_{\tau3} \\
\end{array}
\right)  \nonumber \\
& = &
\left(
\begin{array}{ccc}
1 & 0   & 0 \\
0 & c_{23} & s_{23} \\
0 & -s_{23} & c_{23} \\
\end{array}
\right)
\left(
\begin{array}{ccc}
c_{13} & 0   & s_{13}e^{-i\delta} \\
0 & 1 & 0 \\
-s_{13}e^{i\delta} & 0 & c_{13} \\
\end{array}
\right)
\left(
\begin{array}{ccc}
c_{12} & s_{12}  & 0 \\
-s_{12} & c_{12} & 0 \\
0 & 0 & 1 \\
\end{array}
\right) \\
& &
\times\,
\mathrm{diag}(1,e^{i\frac{\alpha_{21}}{2}},e^{i\frac{\alpha_{31}}{2}}) \, ; \nonumber
\label{Eqn:PCKMmatrix}
\end{eqnarray}
where \(c_{ij}\) and \(s_{ij}\) represent sin\(\theta_{ij}\) and
cos\(\theta_{ij}\), respectively.
The probability for oscillations from flavour $\alpha$ to $\beta$ as
the neutrino propagates in vacuum may be expressed as:
\begin{eqnarray}
  P(\nu_\alpha \rightarrow \nu_\beta) &
  = & \delta_{\alpha\beta} \nonumber \\
  & &  -
  4 \sum_{i>j}^{3} \mathrm{Re}\left[
    U^{*}_{\alpha i} \cdot U_{\beta i} \cdot
   U_{\alpha j} \cdot U^{*}_{\beta j}
 \right]
 \sin^2 \left(\frac{\Delta m^{2}_{ij}}{4}\frac{L}{E} \right)  \\
 & &  +
 2 \sum_{i>j}^{3} \mathrm{Im}\left[
   U^{*}_{\alpha i} \cdot U_{\beta i} \cdot
   U_{\alpha j} \cdot U^{*}_{\beta j}
 \right]
 \sin \left(\frac{\Delta m^{2}_{ij}}{2}\frac{L}{E} \right) \,; \nonumber
\end{eqnarray}
where $\Delta m^2_{ij} \equiv m^2_i - m^2_j$ is the difference of the
squares of the neutrino masses. 
The probability for the oscillation of antineutrinos is the same as
that for neutrinos with $U\to U^*$. This implies that the last line has a sign opposite to that of neutrino oscillations.
The probability is further modified by the interaction of the
(anti-)neutrinos with the matter through which they travel.
Among the three phases, neutrino oscillations depend only on
$\delta$.
CP-invariance violation (CPiV) occurs if $\delta \ne 0^\circ$ and
$\delta \ne 180^\circ$.
The present status of the determination of the mixing parameters is
presented in table \ref{Tbl:oscillation-parameters}.
Each of the three mixing angles have been determined: $\theta_{12}$ is
known with an uncertainty $\sim 2\%$; $\theta_{23}$ is known with a
precision of $\sim 5\%$; and $\theta_{13}$ with a precision of 
$\sim 5\%$.
It is not known whether $\theta_{23}$ is more than $45^\circ$ or less
than $45^\circ$, this is referred to as the ``octant ambiguity''.
The magnitude of the mass-squared splittings are also known, 
$\Delta m^2_{21}$ with an uncertainty of $\sim 3\%$ and 
$\Delta m^2_{32}$ with an uncertainty of $\sim 5\%$.
The sign of $\Delta m^2_{21}$ is known from measurements of the energy
dependence of the oscillation pattern of electron-neutrinos from the
sun.
The sign of $\Delta m^2_{32}$ remains to be determined.
\begin{table}
  \caption{
    Summary of neutrino oscillation parameters \cite{Beringer:1900zz}.
    In the table the abbreviation SK stands for Super-Kamiokande,
    KL for KamLAND, BOREX for Borexino, DB for Daya Bay, and DC for
    Double Chooz.
    Citations for the various experiments may be found in
    \cite{Beringer:1900zz}.
  } 
  \label{Tbl:oscillation-parameters}
\begin{center}
\begin{tabular}{ccc}
\hline \hline
Parameter & Value ($\pm 3\sigma$) & Examples of Experiments\\
\hline
$\sin^2\theta_{12}$ & $0.312^{+0.052}_{-0.047}$ & SK, SNO, KL, BOREX\\
$\sin^2\theta_{23}$ & $0.42^{+0.22}_{-0.08}$ & SK, K2K, MINOS, T2K \\
$\sin^2\theta_{13}$ & $0.0251^{+0.0109}_{-0.0101}$ & T2K, MINOS, DB, DC, RENO \\
$\Delta m_{21}^2$ & $(7.58^{+0.60}_{-0.59}) \times 10^{-5}$~$\rm eV^2$ & SK, SNO, KL, BOREX\\
$|\Delta m_{32}^2|$ & $(2.35^{+0.32}_{-0.29}) \times 10^{-3}$~$\rm eV^2$ & SK, K2K, MINOS, T2K \\
sign of $\Delta m_{32}^2$ & unknown & SK, MINOS, T2K \\
$\delta$ & unknown & SK, MINOS, T2K \\
\hline \hline
\end{tabular}
\end{center}
\end{table}

Of the parameters, the phase $\delta$ has not yet been determined and
the sign of $\Delta m^2_{32}$ is not known.
The ``normal hierarchy'' refers to the case in which $m_3$ is the
heaviest mass state ($\Delta m^2_{32} > 0$); the ``inverted
hierarchy'' is the case in which $m_3$ is the lightest state 
($\Delta m^2_{32} < 0$).
The normal and inverted hierarchies result in measurably different
oscillation probabilities if neutrinos of sufficient energy are caused
to travel an appropriately long distance through the earth.
The determination of the mass hierarchy and $\delta$ is feasible if
the transition 
$\overset{\brabar}{\nu}_\mu \rightarrow \overset{\brabar}{\nu}_e$ (or
$\overset{\brabar}{\nu}_e \rightarrow \overset{\brabar}{\nu}_\mu$) can
be detected with sufficient rate and cleanliness.
The mass hierarchy may also be explored by experiments searching for
neutrinoless double-beta decay, through cosmological observations and
by measuring the energy dependence of electron-anti-neutrinos produced
in nuclear fission.
Whether $\sin^22\theta_{23}$ is maximal ($\theta_{23}=\pi/4$)
or $\theta_{23}$ is in the first octant ($\theta_{23}<\pi/4$)
or in the second octant ($\theta_{23}>\pi/4$) is another open
question.

A number of measurements have been reported that do not fit into the
three-flavour mixing scenario.
These anomalous  measurements can be interpreted as ``hints'' for the
existence of neutrinos other than the three known species.
Such states are referred to as ``sterile'' as they do not interact via
the weak interaction.
To explain the anomalous measurements, new states are introduced with
masses such that the mass-squared difference to the three known states
is $\sim 1$ eV$^2$.
Oscillations between the known states and the sterile states govern
short baseline oscillations such as electron-neutrino appearance in
muon-neutrino beams over baselines of 0.03\,km to 0.5\,km, the deficit
of reactor neutrinos at very short distance and the deficit of
electron neutrinos from radioisotopes.
The evidence for the existence of sterile neutrinos is controversial
and a program to prove or refute the anomalous measurements
is underway in Asia, the Americas and in Europe.

\section{Elements of the future programme}
\label{Sect:FutureElements}

The future programme must determine the parameters of the S$\nu$M and
make the tests necessary to establish it as a precise and
self-consistent description of nature.
With these goals in mind, the elements of the future programme are
identified in this section.

\subsection{Headline measurements}
\label{SubSect:Headlines}

\subsubsection{Completing the picture}

\paragraph*{Searching for CP-invariance violation}

The CP-phase ($\delta$) can only be determined by measuring the rate
of appearance of a neutrino flavour not present in the beam at
source.
The measurement must be made over a range of values of $L/E$ at which
there is significant interference between terms in the oscillation
formalism that arise from $\theta_{12}$, $\theta_{13}$ and
$\theta_{23}$.
In the short term, searches for CP-invariance violation will be made
by measuring the appearance of $\nu_e$ (or $\bar{\nu}_e$) in a
conventional $\nu_\mu$ ($\bar{\nu}_\mu$) beam.
Of the experiments that will take data this decade, only $\NOvA$ and
T2K have the ability to probe CP-invariance violation.

Since the sensitivity of long-baseline experiments arises from the
interference of oscillation modes, sensitivity to $\delta$ arises from
the product  
$\sin\delta \times \sin\theta_{12} \times \sin\theta_{23} \times
\sin\theta_{13}$. 
Therefore, the determination of $\delta$ using long-baseline
experiments requires that measurements of each of the mixing angles
is available.

The proposed next generation long-baseline experiments fall into two
classes: narrow- and wide-band beam experiments.
Narrow-band beams deliver a narrow range of neutrino energies using the 
off-axis beam technique.
Such experiments are able to make a model-independent search for
leptonic CP-invariance violation by measuring the difference between
the rates of neutrino and the anti-neutrino appearance.
In order to reduce the asymmetry introduced by neutrino interactions
with the earth (the matter effect), the low-energy narrow-band beam is
tuned to match a comparatively short baseline.
The Hyper-Kamiokande experiment is an example of such a
project that is sensitive to events at the first oscillation maximum.
A concept to produce a neutrino beam at the European Spallation Source
serving a detector placed at the second oscillation maximum has
recently been proposed.
Wide-band beams generate a broad band of neutrino energies with a
larger mean energy of $\sim 3$\,GeV.
By placing the detector at a distance of more than 1000\,km from the
source it is possible to exploit the broad band of neutrino energies
to study the energy spectrum of the oscillated neutrinos at both the
first and the second oscillation maximum.
By studying the difference between the oscillated neutrino and
anti-neutrino spectrum, such experiments can make model independent
searches for CP-invariance violation.
The Long Baseline Neutrino Experiment (LBNE) proposed in the USA and
the Long Baseline Neutrino Observatory (LBNO) proposal under
development in Europe plan to search for CPiV in this manner.
In addition, by studying the oscillated neutrino-energy spectrum,
both narrow- and wide-band beam experiments are able to probe the
description of CP-invariance violation provided by the S$\nu$M
outlined above.
The two techniques probe for CPiV in different oscillations regimes;
as discussed below, the matter effect is significant in long-baseline
experiments.
Futher, the short- and long-baseline collaborations have converged on
complementary detector technologies; water-Cherenkov detectors and
liquid-argon time-projection chambers respectively.
These factors mean that the measurements that can be made at 
narrow-band, short-baseline beams are complementary to those that can
be made at wide-band, long-baseline beams. 
Systematic uncertainties arising from the modelling of neutrino
interactions are challenging in both short- and long-baseline
experiments.
However, these systematics will be manifested in qualitatively
different ways owing to the different energies and detector techniques
used.
And, as discussed in section \ref{SubSubSect:TestSnuM}, the study of
oscillations as a function of baseline, $L$, and neutrino energy, $E$,
separately will offer the possibility to observe non-standard effects.
Therefore, the combination of results from short- and long-baseline
experiments will be extremely valuable and should be studied further.

To search for CP-invariance violation in a conventional long-baseline
experiment requires the detection of $\nu_e$ ($\bar{\nu}_e$).
Greatly improved signal purity can be achieved using a technique that
can deliver a large flux of electron (anti-)neutrinos since, in this
case, sensitivity to CP-invariance violation requires the detection of
the appearance of $\nu_\mu$ (or $\bar{\nu}_\mu$) which is readily
accomplished through the charged-current production of muons.
In the Neutrino Factory, intense $\nu_e$ ($\bar{\nu}_e$) beams are
produced from the decay of muons confined within a storage ring.
The Neutrino Factory offers the best sensitivity to leptonic
CP-invariance violation and the best precision on the determination of
$\delta$.
The use of high-power cyclotrons to produce neutrinos from the decays
of muons brought to rest in a large detector has recently been
proposed.
Such an experiment has the potential to measure $\delta$ from the $L$
dependence of the oscillated neutrino beam.

\paragraph*{Determining the mass hierarchy}

The large value of $\theta_{13}$ means that the next generation
long-baseline experiments have the potential to determine the mass
hierarchy. 
Neutrinos interact with the earth as they propagate from source to
detector.
All neutrino flavours may undergo elastic scatters with the atomic
electrons.
In the case of the $\nu_e$ and $\bar{\nu}_e$, the charged current makes
a contribution to the elastic-scattering amplitude.
This leads to a ``matter effect'' through which the oscillation rate
of neutrinos differs from that of anti-neutrinos.
The way in which the oscillated, neutrino-appearance energy spectrum
is modified depends on the sign of the large mass splitting ($\Delta
m^2_{32}$).
In the case of long-baseline experiments, good sensitivity to the mass
hierarchy is obtained for small values of $|\Delta m_{32}^2|/E$ and
appropriately long baselines.
LBNE and LBNO seek to determine the mass hierarchy using this
technique.

It is possible to determine the mass hierarchy by measuring with high
precision the shape of the oscillated, neutrino-disappearance energy
spectrum.
Experiments are being developed that will immerse a large detector
capable of measuring the incident neutrino energy with a precision of
$\sim 3\%$ in the large flux of $\bar{\nu}_e$ produced in nuclear
reactors.

Atmospheric neutrinos remain an important probe of neutrino
oscillations and provide a sensitive technique for the determination of
the neutrino mass hierarchy. 
Supernova neutrinos detected in one or several of the existing or
planned large neutrino observatories would allow the determination of
the neutrino mass ordering although uncertainties in the models of
particle production by supernovae make for a challenging analysis.

\paragraph*{Resolving the octant ambiguity}

Disappearance measurements using atmospheric and
accelerator-generated beams are primarily sensitive to $\theta_{23}$ through a
term in the oscillation probability proportional to 
$\sin^2 2 \theta_{23}$.
As a result, such measurements are able to determine 
$|\theta_{23} - 45^\circ|$ but unable to determine the sign of 
$(\theta_{23} - 45^\circ)$.
The ``octant ambiguity'' refers to the fact that the octant in which
the value of $\theta_{23}$ lies is not yet known.
The sign of $(\theta_{23} - 45^\circ)$ can be determined through a
detailed analysis of the $\nu_\mu \rightarrow \nu_e$ oscillation
pattern, where the first-order term in the oscillation probability is
proportional to $\sin^2 \theta_{23}$, thereby giving rise to an
enhancement in the transition if $\theta_{23} > 45^\circ$ relative
to the corresponding value if $\theta_{23} < 45^\circ$ that gives
rise to the same value of $\sin 2 \theta_{23}$. 
In general, the effects of CPiV, the mass hierarchy and $\theta_{23}$
are intimately entangled in $\nu_\mu \rightarrow \nu_e$ oscillations. 
The complementarity afforded by studying the oscillations with
different baselines and energies, along with the anti-neutrino
channel, allows one to vary systematically the impact of each
parameter in the oscillation pattern and extract each of their
values.

\subsubsection{Testing the standard three-neutrino mixing model} 
\label{SubSubSect:TestSnuM}

In the near term, the precision with which the oscillation rates are
known will improve as the present generation of experiments
accumulate data.
Reactor experiments exploiting the $\bar{\nu}_e$ disappearance channel
and operating at baselines of $\sim$1\,km are sensitive to
$\theta_{13}$.
The appearance channels in accelerator-based experiments operating at
baselines larger than $\sim 100$\,km
($\overset{\brabar}{\nu}_{\mu} \rightarrow \overset{\brabar}{\nu}_{e}$)
are sensitive both to $\theta_{23}$ and to $\theta_{13}$, while the
disappearance channel depends mainly on $\theta_{23}$.
Therefore, it is possible to check the consistency of the S$\nu$M by
comparing the results of the reactor and long-baseline experiments.
Assuming consistency between the results, constraints may be placed on
the value of $\delta$.
A separate constraint on $\delta$ may be derived from the
comparison of oscillation measurements using $\nu_\mu$ and
$\bar{\nu}_\mu$ beams.
A check on the consistency of the S$\nu$M will then be afforded
by the comparison of the allowed range of $\delta$ yielded by the two
techniques.
Taken together, the full data sets from reactor and long-baseline
experiments will place stronger constraints on the mixing matrix than
can be derived from any single experiment on its own.

Such complementarity in the oscillation-measurement programme will
continue to be essential beyond the present generation of
experiments.
To confirm the S$\nu$M as the correct description of nature, or to
establish the existence of entirely new phenomena, will require that
consistency checks of sufficient precision be carried out.
For the future accelerator-based neutrino-oscillation programme, this
puts a premium on the study of neutrino and anti-neutrino oscillations,
the measurement of the energy dependence of the oscillations and the
verification that the S$\nu$M is able to give a consistent, detailed
and precise description of at least two experiments at substantially
different baselines. 

It was noted in section \ref{Sect:Status} that there are a small
number of measurements that are not readily described by the S$\nu$M.
It is possible to interpret these results by postulating the existence
of additional ``sterile'' neutrino states that are electro-weak
singlets and that have masses such that the effective mass-squared
splitting with the three known flavours of neutrino ($\Delta m^2$) is
$\sim$1\,eV$^2$.
To date it has not been possible to give a satisfactory,
self-consistent description of all of the anomalous measurements.

Conceptually, the search for sterile neutrinos can be separated into
experiments which seek to confirm or refute the existence of
oscillations with a frequency corresponding to 
$\Delta m^2 \sim 1$\,eV$^2$ and generic searches which do not target
a particular mass scale.
Reactor- and radio-isotope-source based experiments are sufficient to
study $\overset{\brabar}{\nu}_{e}$ disappearance channels at
oscillation frequencies corresponding to $\Delta m^2 \sim 1$\,eV$^2$.
Accelerator-based experiments will be required to test the
sterile-neutrino interpretation of the observed excesses in the 
$\overset{\brabar}{\nu}_{\mu} \rightarrow \overset{\brabar}{\nu}_e$
appearance and the $\overset{\brabar}{\nu}_{\mu}$ disappearance
channels.
Generic searches for sterile neutrino states should be developed in
synergy with the neutrino-nucleus scattering programme that is
required to support the future long-baseline neutrino-oscillation
programme.
Such a cross-section measurement programme will also be able to set
limits on non-standard phenomena in neutrino-scattering.

\subsection{Experimental programme required to deliver headline measurements}
\label{SubSect:SupportingProgramme}

The long-baseline accelerator-based neutrino-oscillation experiments
will rely on intense neutrino sources and very massive detectors to
reduce statistical uncertainties.
For the experiments to realize their potential, the systematic uncertainties 
will need to be controlled such that they are
always commensurate with the statistical precision.
A concurrent programme of neutrino-nucleus scattering and
hadro-production measurement is therefore required.

\subsubsection{Neutrino Interaction Measurements}

Solar and reactor neutrino experiments operating at very low neutrino
energies (several MeV) and scattering experiments at very high
energies (100's of GeV), enjoy very precise knowledge of their
respective neutrino cross sections (at the few-percent level).
In the neutrino energy range of interest to accelerator based
oscillation experiments, 0.2\,GeV--5\,GeV, the inclusive 
$\overset{\brabar}{\nu}_{\mu}N$ cross section is known with precision
between 5\% and 10\%, and the differential cross sections have
uncertainties in the 30\%--50\% range.
Further, the $\overset{\brabar}{\nu}_{e}N$ cross section has not been
measured in the energy range of interest.

The phenomenology of neutrino interactions in the few-GeV energy
region is a complex combination of quasi-elastic scattering, resonance
production and deep inelastic scattering processes.
Each of these processes has its own model and associated
uncertainties.
The models are based on neutrino cross section measurements that have
substantial uncertainties, are not always consistent from one
measurement to the next and which are often in conflict with
theoretical predictions.
Furthermore, current and future experiments will have to contend with
significant nuclear effects which modify the composition and spectra
of the hadronic final state as well as the observed
neutrino-scattering rate.
Recently, there has been a new appreciation of the importance of
accounting for nuclear effects in the reconstruction of the incoming
neutrino energy and in the calculation of the difference between
neutrino and anti-neutrino cross sections.

Much of the neutrino-scattering data on which phenomenological models
are based are rather old and obtained using target materials not
relevant for modern oscillation experiments.
Taking advantage of new intense sources of neutrinos, modern
experiments have begun to remeasure these neutrino interaction cross
sections, most importantly on nuclear targets relevant to the neutrino
oscillation program. 
Such measurements are performed using near detectors associated with
long-baseline neutrino projects as well as small-scale dedicated
neutrino-scattering experiments. 
While recent data have been instrumental in driving progress, they often
raised more questions than they have been able to answer.
It is clear that additional measurements are needed to complete our
understanding of the few-GeV energy region and the nuclear physics at
play. 
In particular, it has been noted that not all nuclear effects cancel
in the comparison of the event rates at the far detector with those at
the near detector in long-baseline experiments.

A well-considered program of precision neutrino-scattering experiments
in both low- and high-energy regimes on a variety of nuclear targets
for both neutrinos and anti-neutrinos is required, combined with a
dedicated theoretical effort to develop a reliable,
nuclear-physics-based description of neutrino interactions in nuclei
that can be used in neutrino oscillation fits.
The first step in this programme will be to establish a clear set of
goals for the precision with which $\overset{\brabar}{\nu}_{\mu}N$ and
$\overset{\brabar}{\nu}_{e}N$ cross section need to be measured.
In addition, precise measurements on hydrogen or deuterium targets and
the study of electron-neutrino cross sections will be needed to
provide further confidence in our ability to constrain this complex
physics for our future neutrino-oscillation program.

\paragraph*{Hadro-production Measurements}

All current accelerator-based neutrino oscillation experiments use
either an on- or off-axis neutrino beam produced from meson decays. 
The uncertainty in the normalisation and the spectral shape of the
neutrino flux ultimately limits the precision of neutrino-oscillation
and neutrino-interaction measurements.
It is therefore important to develop an improved understanding of the
details of the production mechanisms that give rise to conventional
neutrino beams.
To improve the knowledge of the neutrino flux it is necessary to
determine meson production rates and the underlying meson momentum and
angular distributions.
Such hadro-production measurements can then be combined with detailed
simulations of the optics of the neutrino beam line.
Measurements have already been performed in support of the present
generation of neutrino experiments yielding flux predictions with
uncertainties of $\sim 10\%$. 
Advancing beyond this to support the future accelerator-based
long-baseline neutrino-oscillation program will require
hadro-production experiments pushing to achieve an uncertainty of
$<5\%$ on the neutrino flux.
This will require pion-nucleus scattering as well as
proton-nucleus scattering measurements in order that 
secondary interactions in the target can be simulated with confidence.

\subsection{Required R\&D programme}
\label{SubSect:R&D}

An extensive, coordinated R\&D programme is essential to deliver the
accelerator and detector techniques and systems required to support
the future programme. 
The programme must encompass the development of:
\begin{itemize}
  \item A new generation of high-power, pulsed proton sources;
  \item Intense neutrino sources based on stored muon beams;
  \item Neutrino detector system of unprecedented size and
    granularity;
  \item Magnet systems that can produce an adequate magnetic field
    over a large detector volume in the absence of iron; and
  \item Simulation tools which encapsulate the physics of
    neutrino-nucleus interactions and models of the detector
    response.
\end{itemize}

\paragraph*{Accelerators and beams}

Next generation experiments are based on megawatt-class proton
sources. 
The intensity of such a source introduces many challenges in
accelerator and particle-production-target technology arising from
thermal shock, heating, and radioactivity.
While beam powers of close to 1\,MW will be realised in the current
generation of experiments, multi-MW power will be a new regime. 
Non-destructive beam monitoring technologies that can operate in this
harsh environment are needed. 
In addition to production targets and absorbers that can withstand the
thermal shock, the heat load on components makes the cooling of the
components challenging.
These issues are further exacerbated by such processes as
acidification of air and contamination of, and hydrogen production
from, cooling water.
Such processes give rise to corrosive and/or explosive components in
the environment.   
The high-power environment creates new challenges related to the
maintenance and interchange of highly active components such as
targets and horns.

Developments to realise muon storage rings that would be the basis for
neutrino factories can be divided into two categories.
For nuSTORM, pre-construction R\&D and prototyping is required on the
pion-capture magnet, the large-aperture transport magnets and the
muon-beam instrumentation.
Further R\&D is needed to deliver the high-power target and
pion-capture channel, ionization-cooling system and high-gradient RF
cavities that are required for the Neutrino Factory.

The Panel notes that the promotion of an internationally coordinated
R\&D programme would maximise the impact of the efforts of individual
countries and regions. 

\paragraph*{Detectors}

Future multi-kiloton to megaton-class far detectors have converged on
two technologies: liquid-argon time-projection chambers (LAr) and
water Cherenkov (WC) detectors. 

LAr is the newer technique. Though kiloton-scale single-phase
detectors have already been realised, further developments are required
to prove the dual-phase approach and to construct
multi-kiloton detectors using either the single- or dual-phase
technique.
A separate item is the magnetisation of a large volume detector that
contains no iron.
Magnetising a large LAr detector will allow powerful sign-selection and
momentum-reconstruction techniques that can be applied to data from LAr
detectors, enhancing the ongoing long-baseline efforts.
This capability is also essential for the exploitation of
neutrinos produced by stored-muon beams.

The WC technique has matured over several generations of detectors of
increasing volume.
Significant opportunities remain to enhance photo-sensor and readout
technologies in order to optimise detector performance for a given
cost.
R\&D is required to improve light-collection efficiency, to increase
the effective photo-sensitive area and to develop new amplification
technologies that will significantly reduce the per-channel cost.
A new generation of large-area photo-sensors based on micro-channel
plates is also emerging. 
Other potential enhancements include water-based scintillators that
may give WC detectors new capabilities for detecting and
reconstructing particles with momentum below the Cherenkov threshold.

For near detectors, a wider range of options and  higher levels of
granularity are possible due to the relatively small size of the near
detectors. 
Higher granularity and lower tracking thresholds are in principle desirable to advance studies of
neutrino interactions. 
Options include gaseous detectors such as high-pressure time-projection
chambers and finely segmented scintillator-based tracking detectors. 
In most cases, magnetisation allows sign selection,
momentum-measurement and particle-identification capabilities that are
either beneficial or essential depending on the application.

Again, the Panel notes that an internationally coordinated R\&D
programme would maximise the impact of the efforts of individual
countries and regions.

\paragraph*{Software and computing}

Along with the hardware development for LAr, a vigorous program to
realise reconstruction algorithms that can fully exploit the
information provided by these detectors is needed.  
While reconstruction algorithms have been developed in the context of
existing and past WC detectors, there are opportunities to develop and
improve these algorithms, particularly if new photo-sensor or optical
elements are introduced. 

Neutrino event generators, which simulate the final states produced by
neutrino interactions, are another area where significant development
is needed. 
The reliability and robustness of the underlying models used in these
generators is critical for the projected physics sensitivities in
the next generation experiments to be achieved.
New measurements of neutrino interactions from the current and
upcoming generation of experiments, along with commensurate
developments and improvements in the relevant nuclear theory, will
provide the foundations for these improvements.

Due to the broad range of expertise that will be required, which spans
fields that are traditionally outside the realm of particle physics,
and the fact that this expertise is widely distributed geographically,
international coordination and cooperation are essential for success. 
The Panel plans to take an initiative to promote cooperation and to
share best practice in the software and computing required for present
and future experiments.

\paragraph*{Regional strengths}

Accelerator and beam developments relevant for the next generation of neutrino
experiments are being pursued in Asia, Europe, and 
in the Americas, and particularly at CERN, FNAL and J-PARC, each of which
hosts, or has recently hosted, neutrino oscillation experiments. 
Expertise that is critical to the success of the accelerator R\&D
programme can be found in many other laboratories (e.g. targetry at
RAL, remote handling at RAL and TRIUMF, superconducting magnets at
BNL and FNAL, etc.). 

Likewise, LAr TPC developments are occurring in all three regions. 
At CERN, a large, single-phase, prototyping effort based on the
successful ICARUS project is in progress in the context of WA104.
The WA105 programme seeks to realise a large
($6\times6\times6~\mbox{m}^3$) dual-phase detector and to expose it to
charged-particle beams.
At FNAL, several neutrino-physics collaborations (MicroBooNE and
LAr1ND) are building LAr detectors, while the CAPTAIN and LArIAT R\&D
activities are investigating the performance of LAr in test beams.  
Efforts are also under way in Japan.

For WC detectors, Japan has a strong track record building on the
success of the Kamiokande and Super-Kamiokande experiments.
Antarctic and sea-based experiments such as PINGU, ANTARES and
KM3NET, which detect Cherenkov radiation in sea water or ice, have similar photo-sensor, readout, and calibration requirements
that may offer opportunities for cooperation. 
In addition, significant WC expertise exists in North America from
SNO, Super-Kamiokande, and the R\&D programme for LBNE. 

In all regions, experiences from past experiments using a
wide variety of different technologies are an invaluable repository of
knowledge and expertise that may be applied to the development of the
next generation of neutrino beams and the near and far detectors.
The Panel believes that the development of a mechanism by which this
expertise can be coordinated is essential to maximise the
cost-effectiveness of the R\&D programme.

\subsection{Required theory and phenomenology programmes}
\label{SubSect:TheoryPheno}

Better Monte Carlo (MC) simulation tools are required to allow
systematic uncertainties related to the poor understanding of
neutrino-nucleus-scattering and hadro-production cross sections to be
reduced in line with the statistical power of the oscillation
measurements.
The generators cannot be more precise than the cross-section
measurements on which they are based; i.e. currently with precision
not better than $\sim 20\%$ in the neutrino-energy range of
interest. 
Some limitations arise from the fact that basic parameters such as the
nucleon-$\Delta$ excitation transition matrix elements are poorly
known. 
Such uncertainties can only be reduced through a programme of
measurements.
The use of outdated theoretical models implemented in the generators
is also a source of substantial uncertainty.
As an example, the Fermi Gas (FG) model, which is known to not capture 
relevant physics in neutrino-nucleus interactions, is often used to describe 
the behaviour of nucleons because it is simple to implement. 
These sources of systematic uncertainty can only be improved with a
thorough theoretical programme.
Coordination is necessary to ensure that appropriate models of the
basic physics are implemented consistently across the suite of
simulation tools.

It has become clear that nucleon-nucleon correlations (neglected in
the FG model) must be properly included.
Much useful information can be obtained from ab initio computations of
nucleon-nucleon effects for nuclei such as carbon and oxygen.
Even though such computations are limited by their non-relativistic
character, they put severe constraints on the approximate models that
are widely used.
Another element of the present suite of simulation tools that requires
substantial improvement are the models of the final-state interactions
(FSI).
Typically, these modules rely on simplified cascade approaches.
Electron-scattering data, for which the lepton kinematics is precisely
known, is a rich source of information that can be used to validate
the cascade models since hadronic FSI effects are the same in both
neutrino- and electron-nucleus scattering.

The study of neutrino--nucleon and neutrino--nucleus interactions
benefits from a number of international networking activities
including the ``NuInt'' workshop series and new initiatives such as
the NuSTEC collaboration.
The development of a vigorous phenomenological programme to combine
the world's neutrino-oscillation data is also important.

The ultimate goal of the neutrino oscillation programme is to over-constrain 
the parameters of the standard
three-flavor mixing model in order to test the S$\nu$M paradigm, which
can only be achieved through the careful combination of data from
several different sources.
Once experimental uncertainties are no longer dominated by
statistical errors, global fits that can take into account
correlations between the systematic uncertainties will play a crucial
role.

There are many connections between neutrino oscillations and
observables in other areas of particle physics, particle astrophysics
and cosmology.
Further theoretical and phenomenological exploration of such
correlations is important to develop a deeper understanding of
fundamental physics.
These connections strengthen the case for the pursuit of
neutrino-oscillation experiments.

\section{Towards figures of merit beyond the Standard Neutrino Model}
\label{Sect:FigsOfMerit}

When considering the development of the accelerator-based
neutrino-oscillation programme beyond the present generation of
experiments it is necessary to articulate clearly the goals of the
programme and the figures of merit against which proposed
contributions to the programme can be judged.
The S$\nu$M postulates the existence of three neutrino-mass
eigenstates, linear-combinations of which couple to the charged
leptons and the $W$-boson as prescribed by the Standard Model.
The same states also couple to one-another and to the $Z$-boson and,
except for the feeble gravitational interactions, do not directly
interact in any other way. 
This picture, while sufficient to explain almost all neutrino data, is
yet to be tested experimentally in any non-trivial way.
Large deviations from the three-flavour paradigm are allowed, for
example:
\begin{itemize}
  \item There may be more than three neutrino-mass eigenstates. 
    Theoretically, the number of neutrino types need not be the
    same as the number of charged-lepton or quark flavours. 
    New, sterile-neutrino degrees of freedom are allowed; the
    properties of these new states are constrained only very weakly by
    data.
    Sterile neutrinos can only be ``observed'' via their mixing with
    the active neutrinos and, if their masses are small enough, will
    only manifest themselves in neutrino-oscillation experiments. 
    Broadly speaking, evidence for sterile neutrinos can be searched
    for in two ways:
    \begin{enumerate}
      \item New mass eigenstates imply the existence of new
        mass-squared differences and hence new neutrino-oscillation
        lengths, the existence of which can only be probed in
        oscillation experiments. 
        The short-baseline anomalies are often interpreted as evidence
        for new neutrino-oscillation lengths.  
        An integral component of the neutrino-oscillation programme
        must therefore be to perform experiments to establish or
        refute the existence of these new oscillation lengths for
        amplitudes large enough to accommodate the short-baseline
        anomalies.
        The confirmation of the existence of more than two oscillation
        lengths would revolutionise particle physics and open up a new
        window for exploration of the neutrino sector.  

        Even if it becomes clear that the short-baseline anomalies are
        not a consequence of new oscillation lengths, the hypothesis
        that there are light sterile-neutrinos remains intriguing. 
        In the absence of any experimental hints, however, it is
        important to evaluate whether there are parts of the
        mass--mixing parameter space which are theoretically
        preferred. 
        For example, light sterile neutrinos may be a ``side-effect''
        of the elusive mechanism behind the very light neutrino
        masses.
        If neutrinos are Majorana fermions and the new mostly sterile
        mass states are heavier than the mostly active ones, then it
        may be expected that the new mixing angles $\theta_{\rm new}$
        are related to the new masses $M_{\rm new}$;
        $\theta_{\rm new}^2\sim m_{\rm light}/M_{\rm new}$, where
        $m_{\rm light}$ are the mostly-active-neutrino masses. 
        This relation provides a concrete target for sterile-neutrino
        searches. 
        Another example is the possible evidence for new ``neutrino''
        states from cosmology, which is currently in a state of flux.
        If these are indeed neutrinos, their properties will be
        constrained by cosmological data (including, perhaps, some
        currently-unaccounted-for interactions) and searching for
        them in accelerator-based experiments will be of the utmost
        importance.
      \item The neutrino-mixing matrix, as defined within the
        S$\nu$M, is unitary only if there are no new neutrino states.
        New neutrino states imply that the ``correct'' mixing matrix
        is larger than three-by-three, so any sub-matrix is not
        constrained to be unitary. 
        If the new oscillation lengths are too short, the presence of
        new states can only be revealed indirectly via detailed
        unitarity tests, which require precision measurements of
        neutrino oscillations in a variety of channels, including both
        appearance and disappearance modes;
    \end{enumerate}
  \item Neutrinos may participate in new, ``weaker-than-weak'',
    interactions. 
    New neutrino interactions of the neutral-current type will
    mediate, at leading order, non-standard matter effects. 
    These can only be directly probed in long-baseline
    neutrino-oscillation experiments.
    Usually, complete models for non-standard neutrino--matter
    interactions are also constrained by other types of precision
    experiments and by searches for rare processes involving charged
    leptons.  
    However, there are scenarios, at least at the ``existence-proof''
    level, that can only be constrained by neutrino oscillations; and
  \item  Neutrino propagation may deviate from standard expectations.
    Neutrino oscillations require the existence of a
    macroscopically-coherent source of neutrinos and rely heavily on
    the fact that neutrinos obey the dispersion relation of
    fundamental fermions expected by special relativity. 
    For this reason, neutrino oscillations are very sensitive and
    quite unique probes of Lorentz invariance and allow for searches
    for new sources of quantum-mechanical decoherence from
    hypothetical ``neutrino--vacuum'' interactions. 
    The tiny neutrino masses also translate into some very stringent
    tests of the CPT-theorem, including the prediction that fermions
    and anti-fermions have exactly the same mass.   
\end{itemize}

Regardless of whether new degrees of freedom, interactions, or
fundamental principles are revealed, neutrino oscillation experiments
will provide very precise measurements of the neutrino mass-squared
differences and the elements of the leptonic mixing matrix. 
Neutrino masses and mixing angles, in turn, are basic pieces of the
long-standing flavour puzzle---where do the observed values of fermion
masses and mixing angles ``come from''?  
There are many models and ideas proposed and discussed at length in
the literature but we are far from a satisfactory answer to this
question.

In the context of the Panel's initial report, it is more instructive
to identify specific targets for the measurement-precision required of
the future programme.
While there are no unambiguous answers, a couple of options that can
serve as ``robust'' targets, more or less independent of the 
would-be flavour model, have been identified:
\begin{itemize}
  \item Predictions from flavour models often translate into algebraic
    relations among the different mixing parameters
    (e.g. $\theta_{23}=\pi/4+\sqrt{2}\theta_{13}\cos\delta$). 
    In order for these relations to be tested robustly, the
    precision of the different components of the relation should be
    known with similar precision. 
    The smallest known mixing angle, $\theta_{13}$, is of order a few
    percent, indicating that, in order to test different relations
    among mixing angles and the CP phase, $\delta$, it is important to
    measure all mixing angles with a precision comparable to the
    precision with which $\theta_{13}$ will be known; and 
  \item Some flavour models relate mixing angles, or their deviations
    from special values, to small parameters in the theory. 
    In the neutrino sector there are a few small parameters, including
    $\sin^2\theta_{13}$, $\cos2\theta_{23}$ and the ratio of the
    mass-squared differences ($\Delta m^2_{21}/\Delta m^2_{31}$). 
    The ratio of the mass-squared differences, for example, provides
    successive targets for precision: 
    $\sqrt{\Delta m^2_{21}/|\Delta m^2_{31}|}\sim 17\%$; 
    $\Delta m^2_{21}/|\Delta m^2_{31}|\sim 3\%$; and 
    $(\Delta m^2_{21}/\Delta m^2_{31})^2\sim0.1\%$.
\end{itemize}

Detailed exploration of the neutrino sector, combined with searches
for new degrees of freedom at the LHC, the search for rare or
forbidden processes and the precision measurement of the global
properties of the Universe, to name a few, along with the necessary
theoretical work to ``tie'' everything together, will all be required
to construct a more satisfactory description of Nature at the smallest
distance scales.
Regardless of the nature of the new physics, accelerator-based
neutrino-oscillation experiments will play a unique and fundamental
role.

\section{Opportunities}
\label{Sect:Opportunities}

To carry out the ambitious experimental programme by which all of the
properties of the neutrino are determined will require coordinated
investment in the necessary facilities; including long- and
short-baseline oscillation experiments and the infrastructures
necessary to determine, with the requisite precision, neutrino flux
and neutrino-nucleus scattering cross sections.

\subsection{The approved program}

The long-baseline oscillation programme is being taken forward by the
T2K experiment in Japan and by the MINOS+ and $\NOvA$ experiments in the
US.
Over the next three years, MINOS+ will collect a data sample $\sim 60$
times that of MINOS with a neutrino-energy spectrum peaked between
4\,GeV and 10\,GeV.
The large event rate and the relatively broad energy spectrum will
allow MINOS+ to search for evidence of oscillation phenomena not
described by the S$\nu$M. 
T2K and $\NOvA$ exploit the kinematics of pion decay to obtain a
narrow-band neutrino beam tuned to the separation between the source
and the far detector.
In the case of T2K, the beam energy is matched to the 295\,km distance
between J-PARC and the 22.5\,kT fiducial mass Super-Kamiokande water
Cherenkov detector.
T2K has recently observed electron-neutrino appearance in a
conventional muon-neutrino beam.
Data taking over the next five to seven years will allow $\theta_{23}$
to be determined with a precision of $\sim 2$\%.
The direct comparison of the T2K measurement with the $\theta_{13}$
determined through the reactor-neutrino $\bar{\nu}_e$ disappearance
measurement will allow a first search for CPiV to be made with a
sensitivity at the $\sim 1\sigma$ level for 
$-150^\circ \lesssim \delta \lesssim -30^\circ$.  
$\NOvA$ is a 14\,kT liquid-scintillator detector placed at a distance of
810\,km from FNAL and is illuminated with the NuMI beam.
The comparatively long baseline gives $\NOvA$ sensitivity to the mass
hierarchy.
With an exposure of six years in neutrino and anti-neutrino modes,
$\NOvA$ will be able to determine,at the $2\sigma$ level, that 
$\Delta m^2_{31} \lesssim 0$ if $-150^\circ \lesssim \delta \lesssim -10^\circ$ or that 
$\Delta m^2_{31} > 0$ if $20^\circ \lesssim \delta \lesssim 140^\circ$.

Neutrino fluxes are estimated using simulations that exploit
parameterisations of the spectra of hadrons produced in pion- or
proton-nucleus interactions.
Following on from the HARP experiment at CERN and the MIPP experiment
at FNAL, the NA61/SHINE experiment at CERN will provide measurements
in the kinematic range of interest to the present generation of
long-baseline oscillation experiments.
The goal is to allow the absolute flux of the
J-PARC neutrino beam to be estimated with a precision of 
$\approx 5\%$ and the ratio of the flux at the near and far detectors
to be estimated at the $\approx 3\%$ level.
Neutrino-nucleus scattering measurements are being made by the
cross-section dedicated ArgoNeut and MINERvA experiments as well as by
the oscillation experiments MINOS(+), MiniBooNE, $\NOvA$, SciBooNE, T2K
(ND280) and, in the near future, MicroBooNE.
Single- and double-differential $\nu_\mu N$ ($\bar{\nu}_\mu N$)
cross-section measurements with statistical and systematic
uncertainties at the 10\% to 30\% level will be provided.
By exploiting the small electron-neutrino contribution to the
conventional-beam flux, the present generation of experiments will be
able to determine single-differential $\nu_e N$ ($\bar{\nu}_e N$)
cross sections at the 20\% to 50\% level over a limited range of the
parameter space. 

The short-baseline accelerator-based neutrino program is focused on
elucidating the nature of the anomalies observed by the LSND and
MiniBooNE experiments.
These anomalies consist of the original evidence for
$\overline{\nu}_\mu \rightarrow \overline{\nu}_e$ transitions from the
LSND experiment, supported by the equivalent measurements of 
MiniBooNE, and an excess of $\nu_e$-like interactions at low energies
observed in the MiniBooNE experiment.  
The origins of these anomalies need to be understood.
A new experiment, MicroBooNE, will begin taking data with a 170\,Tonne
liquid Argon TPC at the Fermilab Booster Neutrino Beam later this
year. 
MicroBooNE will either confirm or exclude the excess of low-energy
$\nu_e$-like events seen by MiniBooNE and will be able to tell if
these anomalous events are really $\nu_e$ interactions or some other
unexpected type of events producing energetic photons. 
Within the coming four or five years these results are expected to
guide our interpretation of the anomalous MiniBooNE low-energy events
but are not expected to directly address the LSND anomaly.

\subsection{Experimental opportunities: near-future}

To step beyond the approved programme requires the unambiguous
determination of the mass hierarchy and a first attempt to observe
CPiV. 

The Long Baseline Neutrino Experiment (LBNE) will measure $\nu_e$
($\bar{\nu}_e$) appearance in a $\nu_\mu$ ($\bar{\nu}_\mu$) beam using
a 35\,kT liquid-argon time-projection chamber.
The neutrino beam from FNAL which will illuminate the detector will
span a wide range of neutrino energies (a wide-band beam).
The mean of the neutrino-energy distribution matches the baseline of
1300\,km.
For an exposure of ten years LBNE has sensitivity to the mass
hierarchy at the $\sim 5-6\sigma$ level over the full CPiV parameter
space. 
The same exposure will allow the energy spectrum of the
$\nu_e$-appearance signal to be investigated giving sensitivity to 
CPiV at the $\sim 3.5\sigma$ level over 50\% of the CPiV parameter
space.
LBNE is being considered within the US Critical Decision (CD) process
and has received CD0 and CD1 approval.
A budget line for an investment of \$867M in the LBNE programme from
the DOE has been identified.
To deliver the full science programme requires that additional, non-US
partners join the effort.
The Panel recognises that the LBNE collaboration and the FNAL
management are in the process of negotiating non-US contributions to
the programme and welcomes the recent positive developments in
expanding the consortium.

The Tokai to Hyper-Kamiokande (Hyper-K) experiment will study $\nu_e$
($\bar{\nu}_e$) appearance in a $\nu_\mu$ ($\bar{\nu}_\mu$) beam using
a water Cherenkov detector with a fiducial mass of 560\,kT. 
Hyper-K will be illuminated by the J-PARC neutrino beam at a baseline of
295\,km.
Assuming that either the mass hierarchy is known or that it will be
determined using its atmospheric-neutrino data, Hyper-K has
sensitivity to CPiV at the $3\sigma$ level over 76\% of all possible
values of $\delta$.
Hyper-K has been identified as one of the high priority projects on the
MEXT road-map.
The necessary pre-construction R\&D programme is underway.
The Science Council of Japan has identified a planning line at the
level of \textyen80B for the far detector
and \textyen3B for a near detector system.

FNAL has articulated upgrade plans by which the proton-beam power
serving the long-baseline programme will be increased to 1.2\,MW and
ideas to realize multi-MW beams at J-PARC are being discussed.
In addition, the LBNE and Hyper-K collaborations are planning large
detectors with large fiducial mass with the goal of bringing the
statistical uncertainty of the oscillation measurement down to the
per-cent level.

So, for the future long-baseline programme to realise its potential,
the systematic uncertainties related to neutrino flux and
neutrino-nucleus scattering cross sections must be reduced such that
they are always commensurate with the statistical uncertainties.
Since the signal for the headline measurements outlined above is the
appearance of $\nu_e$ ($\bar{\nu}_e$), the accurate determination of
the $\nu_e N$ ($\bar{\nu}_e N$) cross sections over the kinematic
range of interest will become a priority.

Design studies for alternative facilities are also underway.
The LAGUNA-LBNO consortium is carrying out a design study of the Long
Baseline Neutrino Observatory (LBNO) which would illuminate a suite of
detectors in a mine in Finland with a conventional, wide-band beam
from CERN.
The baseline of 2300\,km would allow the experiment to determine the
mass hierarchy at the 5$\sigma$ level for all values of $\delta$ in
a few years of running.
The wide-band beam would allow LBNO to study the neutrino energy
spectrum in order to search for CPiV with a sensitivity at the level
of 3$\sigma$ over $\sim 55\%$ of the CPiV parameter space assuming a
ten-year exposure of a 20\,kT LAr detector with the nominal SPS beam
power of 750\,kW.
A number of other design studies are also under way.  
The ``Cherenkov Detector in Mine Pits'' (CHIPS) projects would
instrument the water collected in disused mine pits which are in the
path of the NuMI or LBNE neutrino beams.
Results that are competitive with the combined T2K/$\NOvA$ sensitivity to
$\delta$ can be obtained with six years of running.
It has been proposed to provide a beam from the European Spallation
Source (ESS) to illuminate a large water Cherenkov detector placed at
the second oscillation maximum.
The ``ESSnuSB'' experiment will benefit from the fact that the rate of
events at the second oscillation maximum depends strongly on
$\delta$.
Finally, it has been proposed to exploit the $\delta$-dependence of
the oscillation as a function of baseline length ($L$) by illuminating
a single large detector with neutrinos produced by muon decay at
rest (the ``Daedalus'' experiment).
The concept calls for three, high-power cyclotrons placed at distances
of 1.5\,km, 8\,km and 20\,km from the detector.

Turning to the short-baseline programme, there is general agreement
that a new and as-yet-to-be-approved experiment is needed finally to
resolve the origin of the LSND anomaly and that the new initiative
should be designed to be definitive. 
Several candidate experiments have been proposed in Europe and in the
US.
The present accelerator-based proposals include using liquid argon
TPCs at two or more baselines at either Fermilab or CERN; mounting
an improved LSND-like experiment at ORNL; or building a dedicated low
energy muon-decay ring with straight sections pointing at a
Neutrino-Factory type detector (nuSTORM).
Although the scope of these proposed initiatives could in principle be
accommodated within a regional budget, the interest in doing the right
experiment is international. 
The choice and execution of the next generation short-baseline
experiment(s) would therefore seem fertile ground for fully
international cooperation.

The contributions to the systematic uncertainties may be broken down
into three broad classes.
There are uncertainties relating to the particular detector
configuration; these uncertainties must be addressed by each
collaboration individually.
Then, there are uncertainties relating to the estimation of the
neutrino flux and uncertainties relating to the
neutrino-nucleus-scattering cross sections.
To manage the systematic uncertainties related to the flux it will be
necessary to consider the hadro-production measurement programme that
may be required to follow on from the NA61/SHINE experiment.
Precise knowledge of the neutrino-nucleus-scattering cross sections
and final-state spectra are required for the reactor, atmospheric,
solar and cosmological neutrino programmes as well as the
accelerator-based programme; the measurement of the cross sections is
therefore an important service that the present and future
accelerator-based neutrino-physics programme must provide.
Neutrino detectors exploit water, liquid argon, liquid scintillator or
iron as the material in which neutrinos are captured.
To develop an understanding of the cross-section phenomenology
sufficient to allow extrapolation using Monte Carlo techniques is
likely to require the measurement of neutrino scattering on additional
materials.
The measurement of $\nu_e N$ ($\bar{\nu}_e N$) cross sections with the
requisite per-cent-level precision will require the development of a
novel neutrino source such as that proposed for nuSTORM.
The development of an appropriate, systematic study of
neutrino-nucleus scattering measurements will become of increasing
importance to the field.

\subsection{New experimental opportunities: long-term}

The objectives of the neutrino programme identified in section
\ref{Sect:Intro} are:
\begin{enumerate}
  \item To determine the properties of the neutrino that are presently
    unknown;
  \item To determine whether the S$\nu$M picture is the whole story;
    and
  \item To measure the properties of the neutrinos with a precision 
    sufficient to allow the physics that explains these properties to
    be understood.
\end{enumerate}
The next generation of experiments will improve the precision with
which the mixing angles and mass-squared differences are known and
determine the mass hierarchy.
With $\delta\in[-180,180^\circ]$, if $|\delta| \lsim 25^\circ$ ($|\delta| \gsim 155^\circ$) a new and
novel technique will be required to continue the search for CPiV.
In any event, a new technique will be required to test the
S$\nu$M and to determine the neutrino-mixing parameters with a
precision sufficient to allow the underlying physics to be
elucidated.
The Panel therefore concludes that a programme of accelerator and
detector R\&D is needed to deliver the technologies required to drive
the field beyond the sensitivity and precision offered by the next
generation of experiments.

The Neutrino Factory, in which intense beams of electron- and
muon-neutrinos are produced from the decay of stored muon beams, has
been shown to offer the ultimate sensitivity to CPiV and precision on
$\delta$.
The charge-to-mass ratio of the muon makes it possible to tune the
stored-muon-beam energy to provide a neutrino beam matched to a
particular choice of baseline, detector technology or to respond to
changes in the understanding of the physics of neutrino oscillations.
The incremental development of the Neutrino Factory has been studied
by both the International Design Study for the Neutrino Factory (the
IDS-NF) and within the US Muon Accelerator Program (MAP) by the Muon
Accelerator Staging Study (MASS).
Each of these studies has identified a staged implementation of the
Neutrino Factory in which each step is capable of delivering first
rate neutrino-physics and of supporting the R\&D necessary to prepare
the next step in the incremental programme.
The first step in the programme is nuSTORM in which a stored muon beam
with a central momentum of 3.8\,GeV/$c$ and a momentum spread of 10\%
will:
\begin{itemize}
  \item Deliver detailed and precise studies of electron- and
    muon-neutrino-nucleus scattering over the energy range required by
    the future long- and short-baseline neutrino oscillation
    programme;
  \item Make exquisitely sensitive searches for sterile neutrinos in
    both appearance and disappearance modes; and
  \item Provide the technology test-bed required to carry-out the R\&D
    critical to the implementation of the next increment in the
    muon-accelerator based particle-physics programme.
\end{itemize}
The development of the nuSTORM ring, together with the instrumentation
required for the $\nu N$-scattering and sterile-neutrino-search
programmes will allow the next step in the development of muon
accelerators for particle physics to be defined.
Just as the Cambridge Electron Accelerator,
built by Harvard and MIT at the end of the '50s, was the first in a
series of electron synchrotrons that culminated in LEP, nuSTORM has
the potential to establish a new technique for particle physics that
can be developed to deliver the high-energy $\nu_e$ ($\bar{\nu}_e$) 
beams required to elucidate the physics of flavour at the Neutrino
Factory.
The development of muon accelerators for particle physics clearly
requires international collaboration.

\section{Initial conclusions and next steps}
\label{Sect:NextSteps}

\subsection{Initial conclusions}
\label{SubSect:InitConc}

Following its initial consultations with the accelerator-based
neutrino-oscillation community at the three regional Town Meetings,
the Panel has drawn the following conclusions:
\begin{enumerate}
\setcounter{enumi}{-1}
  \item The study of the neutrino is the study of new phenomena that
    are not described by the Standard Model.
    Accelerator-based neutrino-oscillation experiments are an
    essential part of the neutrino-physics programme and offer
    exciting and unique insights into the physics of fundamental
    particles.
    The results of the accelerator-based experiments will have
    important consequences for particle astrophysics and cosmology.
    This breadth of impact justifies a far-reaching experimental
    programme;
  \item The accelerator-based programme is vibrant and is
    international in both intellectual interest, engagement and in
    scope;
  \item The optimal exploitation of the present and approved
    experiments will benefit from increased cooperation in the
    development of better models for neutrino-event generators, 
    event-reconstruction algorithms for large-volume liquid-argon 
    time-projection chambers and frameworks for the robust combination 
    of results from different experiments;
  \item LBNE and Hyper-K offer complementary approaches to the search for
    CPiV.
    The Panel welcomes the positive developments in the LBNE and Hyper-K
    approval processes.
    A dedicated, coordinated programme of measurement is required to
    ensure that systematic uncertainties are commensurate with the
    statistical power of these experiments;
  \item Design studies are underway for conventional wide-band beams
    (LAGUNA-LBNO, which will report in 2014) and ESSnuSB.
    Potentially, these proposals offer attractive alternatives to LBNE
    and Hyper-K.
    In addition, the design of a novel experiment (Daedalus) to search
    for CPiV using neutrinos generated by muon decay at rest is being
    studied; 
  \item The Neutrino Factory in which intense electron- and
    muon-neutrino beams are generated from the decay of muons confined
    within a storage ring remains the facility that offers the best
    sensitivity.
    The incremental, or staged, implementation of the facility is
    being studied.
    The Panel recognises nuSTORM as an attractive first step;
  \item The anomalies in neutrino-oscillation measurements that can
    be interpreted as evidence for ``sterile'' neutrinos are being
    investigated energetically through a programme of short-baseline
    experiments.
    The short-baseline programme required to resolve these anomalies
    convincingly must be developed such that it also benefits the
    long-baseline programme.
\end{enumerate}

\subsection{Next steps}
\label{SubSect:NextSteps}

To optimise the discovery potential of the future oscillation
programme, thereby maximising the scientific return on investment,
requires that the international neutrino community has timely access
to a number of complementary, powerful neutrino-beam facilities.
For the programme to reach its full potential requires that the
``headline programme'' be supported by a programme of measurement by 
which the systematic uncertainties can be made commensurate with the
statistical power of the oscillation experiments.
To ensure timely access to the necessary facilities it will be 
necessary to exploit to the full the infrastructures that exist at
CERN, J-PARC and FNAL such that each region makes a unique and
critically-important contribution to the programme.
To maximise the impact on the programme of the expertise, experience,
resources and infrastructure that exists in laboratories and
institutes worldwide requires active coordination.
Therefore, in its second year the Panel will consult with laboratory
Directors, funding-agency representatives, the accelerator-based
neutrino-oscillation community and other stakeholders to:
\begin{itemize}
  \item Develop a road-map for the future accelerator-based
    neutrino-oscillation programme that exploits the ambitions
    articulated at CERN, FNAL and J-PARC and includes the programme of
    measurement and test-beam exposure that will ensure the programme
    is able to realise its potential;
  \item Develop a proposal for a coordinated ``Neutrino RD''
    programme, the accelerator and detector R\&D programme required to
    underpin the next generation of experiments; and 
  \item To explore the opportunities for the international
    collaboration necessary to realise the Neutrino Factory.
\end{itemize}
The Panel's vision is that, taken together, the road-map and Neutrino
RD programme will form the basis of the ``International Neutrino
Programme'' (I$\nu$P) necessary to deliver the measurements required
for the phenomena that explain neutrino oscillations to be
discovered.

The Panel will exploit the XXVI International Conference on Neutrino
Physics and Astrophysics which will take place in Boston in June 2014
to:
\begin{itemize}
  \item Initiate a discussion amongst the international
    accelerator-based neutrino-oscillation community on its emerging
    vision for an International Neutrino Programme; and
  \item Launch a workshop, or a number of workshops, to promote
    cooperation on the topics noted in item 2 section
    \ref{SubSect:InitConc}.
\end{itemize}

\clearpage
\bibliographystyle{99-Styles/utphys}
\bibliography{Concatenated-bibliography}

\providecommand{\href}[2]{#2}\begingroup\raggedright\begin{thebibliography}{1}

\bibitem{Beringer:1900zz}
{\bfseries Particle Data Group} Collaboration, J.~Beringer {\em et al.},
  ``{Review of Particle Physics (RPP)},''
\href{http://dx.doi.org/10.1103/PhysRevD.86.010001}{{\em Phys.Rev.} {\bfseries
  D86} (2012) 010001}.

\bibitem{ICFAnuPanel:Mandate:2013}
{The International Committee on Future Accelerators}, ``{ICFA Neutrino
  Panel}.'' {http://www.fnal.gov/directorate/icfa/neutrino\_panel.html}, 2013.

\bibitem{ICFAnuPanel:ToR:2013}
{The International Committee on Future Accelerators}, ``{ICFA Neutrino Panel:
  terms of reference}.''
  {http://www.fnal.gov/directorate/icfa/files/Terms-Of-Reference.pdf}, 2013.

\bibitem{ICFA:nuPanelWWWSite}
{The ICFA Neutrino Panel}, ``{ICFA Neutrino Panel}.''
  {http://www.fnal.gov/directorate/icfa/}.

\end{thebibliography}\endgroup

\cleardoublepage
\appendix
\section{The ICFA Neutrino Panel}
\label{App:ICFA-nu-Panel}

ICFA established the Neutrino Panel with the mandate
\cite{ICFAnuPanel:Mandate:2013}: 
\begin{quote}
  {\it To promote international cooperation in the development of the
    accelerator-based neutrino-oscillation program and to promote
    international collaboration in the development a neutrino factory
    as a future intense source of neutrinos for particle physics
    experiments.
  }
\end{quote}
The membership of the Panel agreed by ICFA at its meeting in February
2013 is shown in table \ref{Tab:PanelMembers}.
The terms of reference for the panel \cite{ICFAnuPanel:ToR:2013} may
be found on the Panel's WWW site \cite{ICFA:nuPanelWWWSite}.
\begin{table}[h]
  \caption{Membership of the ICFA Neutrino Panel.}
  \label{Tab:PanelMembers}
  \begin{center}
    \begin{tabular}{|l|l|}
      \hline
      {\bf Name}      & {\bf Institution}                      \\
      \hline
      J. Cao          & IHEP/Beijing                           \\
      A. de Gouv\^ea  & Northwestern University                \\
      D. Duchesneau   & CNRS/IN2P3                             \\
      R. Funchal      & University of Sao Paulo                \\
      S. Geer         & Fermi National Laboratory              \\
      S.B. Kim        & Seoul National University              \\
      T. Kobayashi    & KEK                                    \\
      K. Long (chair) & Imperial College London and STFC       \\
      M. Maltoni      & Universidad Automata Madrid            \\
      M. Mezzetto     & University of Padova                   \\
      N. Mondal       & Tata Institute for Fundamental Resarch \\
      M. Shiozawa     & Tokyo University                       \\
      J. Sobczyk      & Wroclaw University                     \\
      H. A. Tanaka    & University of British Columbia and IPP \\
      M. Wascko       & Imperial College London                \\
      G. Zeller       & Fermi National Accelerator Laboratory  \\
      \hline
    \end{tabular}
  \end{center}
\end{table}

\clearpage
\section{Reports on Regional Town Meetings}
\label{App:RegTwnMtg}

\subsection{Asia}
\label{SubApp:AsiaTwn}

The Asian neutrino community meeting took place  
in the afternoon of November 13, 2013 on the last day of the NNN13 workshop
at Kavli IPMU in Kashiwa city, Chiba, JAPAN.  
This meeting was organized
by the Asian panel members (J.~Cao, S.B.~Kim, T.~Kobayasi, N.~Mondal, and M.~Shiozawa)
to collect input from the neutrino community in Asia and to receive reports 
from regional planning efforts.  
There were approximately 40 participants and the meeting program and presentation files have been made 
available on the web at:\\http://indico.ipmu.jp/indico/getFile.py/access?contribId=8\&resId=0\&materialId=slides\&confId=26\\.
The meeting consisted of an introductory talk, a theoretical presentation on neutrino physics,
and a series of talks about the status and planning of neutrino experiments in China, India, Japan, and Korea with 
an emphasis on international accelerator-based neutrino oscillation experiments.
Following these talks there
was an open discussion among the meeting participants. 
Opinions were solicited through the web page in advance  
to collect broad inputs to the meeting and this discussion.  \\

\noindent
{\bf Presentation Summary }
\begin{itemize}
  \item {\bf Introduction} by Takashi Kobayashi (KEK)\\
An introduction to the ICFA neutrino panel, including its objectives, charges and 
procedures, was given and the goals of the town meeting were explained.  
It was noted that the panel would like to carry out a review of:
  \begin{description}
    \item[(a)] The present status of the neutrino oscillation program 
               within Asia and the developments that can be expected 
               on a 4-to-7-year timescale; 
    \item[(b)] The discovery opportunities for which the accelerator-based
               neutrino oscillation program must be optimized over
               a 7-to-25-year timescale; and
    \item[(c)] The measurements and R\&D (including software development) 
               that are required for the near-term
               (4-to-7-year) and medium- to long-term (7-to-25-year) program 
               in order to fulfill that potential.
  \end{description}

  \item {\bf Why Neutrinos?} by Hitoshi Murayama (Berkeley \& Kavli IPMU)\\
Although the theory of the strong, weak, and electromagnetic forces
appears to be complete after the discovery of the Higgs boson, 
there are still many unanswered questions in particle physics.
There are at least five missing pieces in the standard model: 
non-baryonic dark matter, the lightness of the neutrino masses,
dark energy, acausal density fluctuations in the early universe, 
and a mechanism to generate the baryon asymmetry in the universe.
In this context, neutrino and nucleon decay experiments are unique probes of
high energy (up to $O(10^{16})$ GeV) physics beyond the standard model.
In order to explain the observed baryon asymmetry of the universe, for instance, 
new sources of $CP$ violation, like $CP$ violation in neutrinos, are needed.  
Similarly, proton decay searches may shed light on the manner in which particles 
convert to anti-particles, another piece of information 
essential to understand how the baryon asymmetry evolved in the history of our universe.

 \item {\bf Neutrino Program in China} by Jingyu Tang (Institute of High Energy Physics, CAS)\\
The Daya Bay experiment is now running and aims to
achieve a 3\% measurement of $\sin^2\theta_{13}$ by accumulating data over the 
next four to five years. 
Its successor, JUNO, is a next-generation
liquid scintillator detector whose primary physics target is the determination of 
the neutrino mass hierarchy using reactor antineutrinos over a $\sim 50$ km baseline.  
It is also expected to measure $\Delta m^2_{21}$, $\Delta m^2_{32}$, and $\sin^2\theta_{12}$
at the 1\% level or better.
High-precision measurements of supernova neutrinos, geo-neutrinos,
and solar neutrinos are also anticipated.
Additionally, the possibility of a $CP$ measurement with a new neutrino 
beam facility based on muon decay, MOMENT, is under discussion.

  \item {\bf Neutrino Program in India} by Sanjib Kumar Agarwalla (Bhubaneswar)\\
Though the headline experiment of the India-Based Neutrino Observatory (INO)
will be the ICAL neutrino detector,   
underground laboratories for double beta decay and direct dark matter detection 
experiments are available as well.
ICAL will primarily study atmospheric neutrinos and is expected to have 
$2.5\sigma$ sensitivity to the mass hierarchy by itself and 
$3.4\sigma$ sensitivity in combination with T2K, NO$\nu$A, and reactor experiments.  
Further improvements in the ICAL event reconstruction lead to 
enhanced sensitivity of $3\sigma$. 
It was also noted that many Indian institutions are involved
in the FNAL neutrino program, including the MIPP, MINOS+, NO$\nu$A,
and LBNE experiments.

  \item {\bf Neutrino Program in Korea} by Kyung Kwang Joo (Chonnam National University)\\
The short-baseline reactor neutrino experiment, RENO, is expected
to improve its precision on its measurement of $\sin^2\theta_{13}$
to $\sim 5\%$ over the next five years.
A longer baseline ($\sim 50$ km) reactor experiment, RENO-50,
is being pursued to perform high-precision measurements of
$\Delta m^2_{21}$, $\Delta m^2_{32}$, and $\sin^2\theta_{12}$,
and to determine the mass hierarchy.
RENO-50 will also be capable of observing neutrinos
from supernova, the Earth's interior, the sun, and J-PARC.  
Though a search for neutrinoless double beta decay is also 
within the scope of the detector, a dedicated double beta decay experiment, AMoRE,
is being planned for a ten year run. 
Construction of a short-baseline neutrino oscillation experiment to study 
the reactor neutrino anomaly is also underway.

  \item {\bf Neutrino Program in Japan} by Tsuyoshi Nakaya (Kyoto)\\
With its approved $7.8 \times 10^{21}$ POT, T2K by itself 
has a chance to exclude $\sin\delta=0$ with 
an expected significance of $\sim 2 \sigma$ if, as its latest data suggest, $\sin\delta=-1$.
In that scenario the mass hierarchy could also be
determined by the combination of measurements at T2K and NO$\nu$A.
Additionally, the precision on its measurement of $\sin^2\theta_{23}$ will be 0.045 (2.6$^\circ$)
assuming $\sin^2\theta_{23}=0.5$.
High statistics studies of atmospheric neutrinos at Super-K will have
$\sim2\sigma$ mass hierarchy determination power
and $\sim 2\sigma$ sensitivity to the $\theta_{23}$ octant
if $\sin^2\theta_{23}=0.6$.
Sterile neutrino searches at KamLAND and the J-PARC/MLF (P56)
are also in preparation. 

On a timescale of $\sim 25$ years
the next-generation underground water
Cherenkov detector, Hyper-Kamiokande (Hyper-K), is being proposed
both to serve as the far detector for a long-baseline 
neutrino oscillation experiment using an upgraded J-PARC 
neutrino beam and as a detector capable of
observing proton decay, atmospheric neutrinos,
and astrophysical neutrinos.
Hyper-K is expected to measure the $CP$ phase with $10-20^\circ$ precision 
and can establish $CP$ violation with $>3\sigma$ significance for $74\%$ of the $\delta$ parameter space.
The significance of the mass hierarchy determination is
expected to reach $3\sigma$ or more using high statistics data from the J-PARC neutrino beam and atmospheric neutrinos.
Additionally, if $\mbox{sin}^{2} 2\theta_{23}<0.995$, the $\theta_{23}$ octant can be resolved at $>2\sigma$.
Owing to its factor of 25 increase in fiducial volume, 
Hyper-K has sensitivity to nucleon decays exceeding what has been 
achieved at Super-K by an order of magnitude or more. 
Finally, discussions are underway within the community
concerning ideas for new facilities to achieve a multi-MW neutrino beam, 
for the development of advanced neutrino detectors such as a large scale liquid Argon TPC,
and for upgrades to KamLAND that will extend its sensitivity to $0\nu\beta\beta$ decay 
into the inverted hierarchy region and beyond.\\

\end{itemize}

\noindent
{\bf Discussion Summary}

\begin {figure}[htb]
\begin{center}
\includegraphics[width=0.85\textwidth] {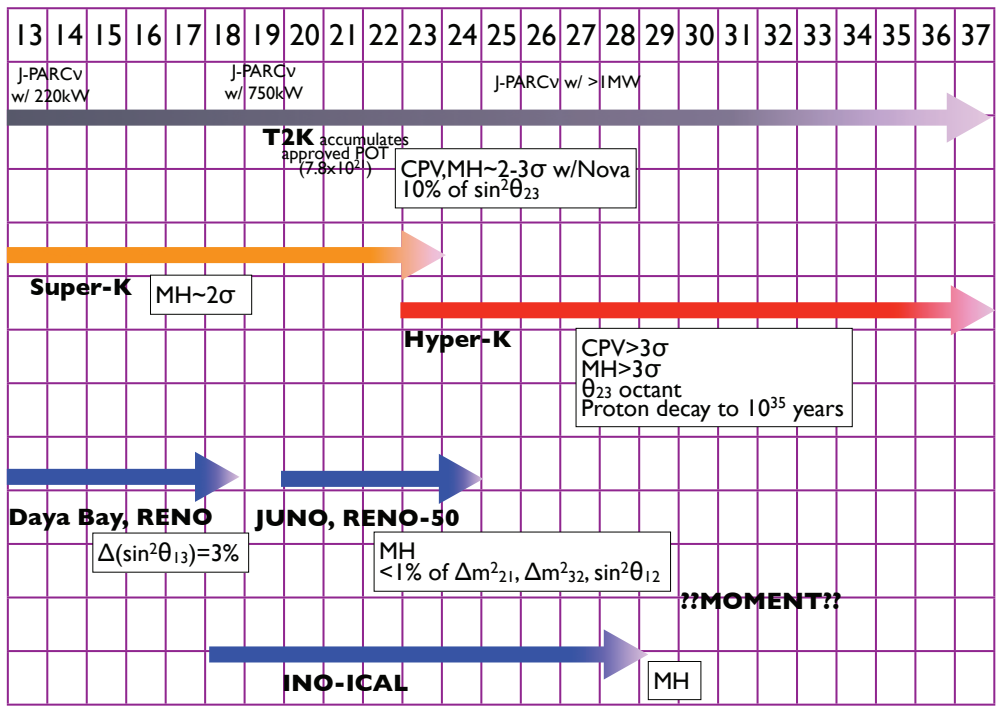}
\caption {Working timeline of neutrino oscillation experiments in Asia and their expected sensitivities.}
\label {fig:asian-timeline}
\end{center}
\end {figure}

Figure~\ref{fig:asian-timeline} summarizes 
a conceptual timeline of both running and planned experiments
with (optimistic) estimates of expected measurement sensitivities. 
T2K is now entering an era of $CP$ violation studies, which 
will be significantly expanded at Hyper-K with the upgraded 
J-PARC neutrino beam.
Indeed, Hyper-K's test of $CP$ violation in neutrinos 
represents a significant opportunity for discovery over 
the next 7-to-25-years.  
The effects of matter on neutrino oscillations 
mimicing $CP$ violation 
are relatively small 
for $\sim 600$ MeV neutrinos 
from J-PARC over the 300 km Hyper-K baseline.
Using well established water Cherenkov technology
Hyper-K consequently offers an experimentally clean and promising 
way to approach the question of neutrino $CP$ violation. 
Moreover, this technology is the only known realistic detector option 
that can probe proton lifetimes of the order of $10^{35}$ years.
High statistics measurements of atmospheric neutrinos at 
Hyper-K and ICAL, as well as measurements of reactor neutrinos over 
moderate baselines at JUNO and RENO-50, each have the potential 
to determine the mass hierarchy with $>3\sigma$ significance. 
These measurements, in conjunction with high precision measurements of 
the neutrino mixing and mass parameters by these projects, 
highlight the strength and versatility of the neutrino program in Asia.

In order to achieve these goals, several essential R\&D studies were 
discussed at the meeting. 
Improvements on the neutrino detection technology are necessary
in the following areas:
the development of fast, high quantum efficiency photo-sensors 
and gadolinium loading technology for large water Cherenkov detectors,
the development of advanced liquid scintillators for scintillator experiments, 
and the development of other advanced detector technologies such as the 
liquid argon TPC.
To achieve a high intensity neutrino beam, upgrades 
to the J-PARC beamline to allow for 750 kW and multi-MW operation 
must be developed and as a part of that program research into 
high power targetry is needed. 
Additionally, both the development of a 15 MW proton driver 
and advances in muon transportation are essential for the success of the 
MOMENT concept.  Central to all upcoming measurements 
is the reduction of systematic errors and accordingly, 
improved hadron production and cross section measurements are critical.

Finally, all participants agreed that
the Asian program can be successful and strengthened by
the mutual support and participation.

\clearpage
\subsection{The Americas}
\label{SubApp:AmericasTwn}

The ICFA Neutrino Panel held a two-day Americas Town Meeting at
Fermilab, 23--24 January 2014. 
The meeting was well attended with approximately 90 participants from
the experimental and theoretical neutrino communities, and with
representatives  from the U.S., Canada, Mexico, and South America. 
There were also a few participants from Asia and Europe. 
The meeting agenda included 25 talks that explored the themes:
\begin{enumerate}
  \item Present issues;
  \item Regional study summaries and perspectives from the Americas;
  \item International collaboration: experience from present
    experiments;
  \item Opportunities for international collaboration. 
\end{enumerate}
In addition, there were three round table discussions that were
moderated by distinguished members from the neutrino community. 
The round tables explored the questions: 
\begin{enumerate}
  \item What coordination is necessary for planning the short-baseline
    neutrino program? 
  \item How to increase collaboration across the Americas?  
  \item What do we need to do to further promote the long-baseline
    science case?
\end{enumerate}

Discussion was encouraged throughout the meeting. 
The discussion associated with the presentations, together with the
presentations themselves, and the round table discussions, revealed a
great deal of consensus within the community about directions and
priorities for the global neutrino program, as well as many
suggestions for areas where increased global coordination might be
fruitful.

\subsubsection{General consensus}

\begin{enumerate}
  \item The presently running, or soon-to-be running,
    accelerator-based neutrino oscillation experiments will advance
    our knowledge of the properties of the neutrino by further
    constraining (or measuring) the $\theta_{23}$ octant, the mass
    hierarchy, and the CP phase /delta, and may also begin to
    elucidate the nature of the excess of the MiniBooNE low-energy
    neutrino events.
  \item Beyond the present experiments, the highest priority for the
    future accelerator-based neutrino program is a new massive
    deep-underground long-baseline experiment. 
    The candidate experiments are LBNE, LBNO, and Hyper-K. 
    There has been significant recent progress in exploring the
    possibility of merging the LBNE and LBNO Collaborations. 
    A new long-baseline possibility using the future European
    Spallation Source is also being explored.
  \item One or more new accelerator-based short-baseline experiments
    are needed if we want to conclusively resolve the origin of the
    LSND and MiniBooNE anomalies.
    The new experiment(s) should be designed to be definitive.
  \item New experiments to measure hadroproduction and neutrino and
    antineutrino cross-sections may be needed to enable the next
    generation of neutrino oscillation experiments to achieve their
    full potential. 
    This deserves further consideration. 
\end{enumerate}

\subsubsection{Increased global coordination}

The discussions generated a number of suggestions for areas which
might benefit from increased international and/or global
coordination. 
The following is an unfiltered list:
\begin{itemize}
  \item  Support for a CTEQ-like group to develop a global
  understanding of neutrino cross-section measurements and the
  associated tools needed for Monte Carlo event generators;
  \item A forum, meeting series, or other mechanism by which
    laboratories can disclose and discuss their capabilities in the
    context of future neutrino projects and R\&D;
  \item The organization of neutrino workshops and schools in Latin
    America;
  \item Early planning of new neutrino beam facilities, leading to a
    more global approach to constructing new facilities and R\&D for
    subsequent upgrades (LARP-like R\&D activities);
  \item Support for a globally coordinated CERN RD-like neutrino
    detector R\&D program;
  \item Support for an international/global effort on liquid argon
    neutrino event-reconstruction;
  \item Planning the program of hadroproduction and neutrino
    cross-section experiments needed to fully exploit future
    long-baseline experiments; and
  \item R\&D on targetry and horns.
\end{itemize}

\clearpage
\subsection{Europe}
\label{SubApp:EuropeTwn}

The meeting took place from January 8\textsuperscript{th} to 10\textsuperscript{th} 
2014 in Paris at the University of Paris Diderot.

The number of participants was 100.

The link to the web site is:

{\emph{http://www.apc.univ-paris7.fr/APC/Conferences/ICFA\_Neutrino\_European\_Meeting\_2014/Home.html}}

\subsubsection{\bf Neutrino facilities discussed at the meeting:} 

{\bf ESSnuSB:} (contribution to Snowmass white paper Sept. 2013)

\leftskip=18pt
The project Concentrates on second oscillation maximum with low neutrino energy around 0.5 
GeV. When coupled to a 440 kton WC detector (Memphys) in a mine at 540 km  it gives a competitive 
CP violation search potential.
The proton beam power  is about a factor 5-6 higher than any other planned proton driver 
for a neutrino beam. Following the ESS schedule, the proton beam will be ready in 2019-2022

A ESSnuSB Design Study will be prepared and submitted to EU (HORIZON2020) in September 
2014.

\vspace{10pt}
\leftskip=-18pt
{\bf NuStorm:} (EOI at CERN May 2013 and proposal to Fermilab July 2013)

\leftskip=18pt
It is a muon based neutrino beam project with nue and numubar of about 2.5-3  GeV. 
This project can add significantly to our knowledge of neutrino interactions and cross sections, 
particularly for nue, since it can provide neutrino beams with fluxes known to 
\texttt{<}1\%.

It offers good capabilities for study of sterile neutrinos with an accelerator beam with 
high sensitivity.
A workforce is setup for nuSTORM to join CENF (CERN neutrino platform activity).
The detector technology can profit for the foreseen R\&D activities in this framework.

They seek to establish a 2-year programme to deliver a Technical Design Report (2016?).

The timescale of this facility is not well defined but the major R\&D required 
and the accelerator modification (new proton driver) would suggest a realisation 
beyond 2020. 

\vspace{10pt}
\leftskip=-18pt
{\bf LBNO and CN2PY:} (EOI at CERN June 2012, Laguna-LBNO FP7 Design study since 2011)

\leftskip=18pt
This is the Long Baseline 2300 km from CERN to Finland (Pyhasalmi) with a 700 kW neutrino beam of 1-10 
GeV.

In a first stahe, a 20 kton Liquid Argon double phase detector is foreseen (run both nu and 
anti-nu  for10-12 years) followed by a 70 kton in the second stage.

They claim to be the only facility to guarantee a mass hierarchy determination at 5 
sigma level after few years running

LBNO has complementarity to HK by providing MH and measuring CP in a different 
way using L/E and the 2nd oscillation maximum

The European design study nearing completion should provide a very detailed costing 
by June 2014. Technical and engineering study work is done is a very serious and 
detailed way.

An R\&D program is developed around the CERN neutrino platform for the double phase 
LAr TP: WA105=\texttt{>} goal is to build a large scale prototype from 2014 to 
2016 and take data in 2017

The funding status of the experiment was not discussed in the meeting however the 
full costing of the project is expected in June.

The situation concerning the Finnish government's interest in the project was not 
reviewed during the meeting.  It is, however, important to note that this issue 
would require additional discussion when the Laguna-LBNO project delivers a report 
with the full cost estimate.

\vspace{13pt}
\leftskip=0pt
\subsubsection{\bf European involvements in LBNE and HyperK:}

LBNE: 

\leftskip=18pt
The project overview was presented by M. Diwan with the US schedule

Europe in LBNE (A. Weber):  10 UK and 8 Italian groups are presented as collaborators

\textit{UK Areas of interest/activities}:  DAQ, 35 t prototype Operation, HV monitoring 
(2014-2017), R\&D TPC components, reconstruction Software and neutrino generators

\textit{Italy Areas of interest/activities}:  Original developer of LAr TPC technology, 
precise scope being discussed, push technology: R\&D and experiment, includes WA104 
@ CERN

LBNO-LBNE Discussions are going on: joint physics task force, common R\&D centred 
around WA105 at CERN (2014-2017) =\texttt{>} comparison of single and double phase, 
 prototyping  common hardware like membrane cryostat

It should be noted that the project is funded in order to cover a 10 kton liquid 
argon detector on surface without a near  detector

Sensitivities are computed with 1\% systematic errors, which are considered too 
optimistic by the community (especially in the absence of a near detector ...)

\vspace{10pt}
\leftskip=0pt
Hyper-K:

\leftskip=18pt
The overview was presented by F. Di Lodovico.

Natural collaborative work emerges from previous European participation in T2K 
=\texttt{>} UK groups showing significant interest.

Ideas to look at improved ND280, at a new near detector at 2 km (WC). 

There will be an LOI submitted to J-PARC in April 2014.

The UK proposal to STCF will be submitted in May 2014

\leftskip=18pt
The areas of interest are: DAQ, electronics, calibration strategy, photodetector studies 
for both Hyper-K and new near detector.

Required R\&D should take place from now to 2017 with a prototyping phase in 2016-2017.

A decision on HK will be taken in 2016. However there could be some ``anticorrelation'' 
with possible strong Japan involvement in ILC

The overall cost is estimated to be 800M\$ (detector only, it does not include 
beam upgrades)

HK should get sensitivity to resolve MH by adding atmospherics to beam data.

\vspace{10pt}
\leftskip=0pt
\subsubsection{\bf Detector and Accelerator R\&D in Europe: }

- A broad range of activities have been presented on the main detector technologies 
that can be used for future neutrino accelerator projects: water Cherenkov R\&D, 
Large liquid Argon TPC, liquid scintillator, magnetised iron-scintillator detector, 
large air core and iron spectrometers. 

- Clearly the needed expertise exists in Europe for all the different technologies. 
  R\&D should be pursued to allow testing large scale prototypes to go beyond the 
different Design Studies. 

In addition an active work is being done in various aspects of accelerators to produce neutrino 
beams. 

The MICE experiment for R\&D towards a Neutrino Factory and a Muon Collider. 

Technological developments for multi-GeV MW-class proton drivers:  CERN, ISIS and 
ESS are three European laboratories which could host a MW-class proton driver for 
a neutrino facility, profiting from existing machines or planned projects.

Target R\&D: materials studies, conceptual designs, Prototyping, Heating/cooling 
tests, beam experiments are covered.

\vspace{0pt}
\leftskip=0pt
\subsubsection{\bf The CERN Neutrino Platform (CENF)}

The development of the new neutrino platform at CERN is a real advantage for the 
community. Indeed it is very important that CERN is becoming involved in neutrino 
physics since it enhances the chances to develop a consistent European neutrino 
strategy.

CERN offers a platform for:

\leftskip=54pt
- Neutrino detectors R\&D (2014-2018)

- Logistics and test beam  infrastructures (2014-2016)

 - Design of a possible neutrino beam.(2012-2014) for an eventual construction in 2015-2018. 
The cost is estimated to be 80-100MCHF.

\leftskip=0pt
- CERN created a budget line for neutrino projects in the Medium Term Plan. It 
will support this platform in an active way and will help WA104, WA105 and others 
in this initial phase. 

- Both projects are finalizing their MOU and are preparing detailed plans.  Approval 
is for R\&D, but no neutrino beam has yet been granted.

- CERN planned involvements:

\leftskip=54pt
Construct a large neutrino test area (EHN1 extension) with charged beams capabilities, 
available in 2016 (and compatible with a future neutrino beam);

Continue the detailed studies towards a short baseline neutrino beam at CERN in 
particular for the secondary beam facility;

Assist the EU neutrino community in their long term common plans.

\vspace{10pt}
\leftskip=0pt
WA104:

- ICARUS-NESSIE is now the experiment WA104.

- The T600 will be transported at CERN where it will be overhauled. 

- A new T150 detector will be constructed using the same technology of the T600 
for the TPCs and new solutions for LAr purification, electronics, light collection 
and complemented with a magnetic field.

- It includes R\&D for air core muon detector (ACM), testing in charged beam, tracking 
capabilities of the ACM detector with high energy muon penetrating LAr-TPC.

\vspace{10pt}
WA105:

- The project is to build a 6x6x6 m\textsuperscript{3} double phase LAr TPC to validate the LBNO 
far detector proposal technology.

- The demonstrator will be exposed to charged hadrons beam in the North Area.

- The demonstrator will be constructed with all the techniques developed in LAGUNA-LBNO 
and needed for the affordable implementation of the far underground detector. It 
will represent a milestone for future long-baseline programs.

- The TDR and MoU are under preparation.

\subsubsection{\bf Conclusions:}

During the meeting two long baseline European projects have been discussed and detailed. One is the ESSnuSB for which a design study will be prepared and submitted to EU (HORIZON2020) and the other one is the Laguna-LBNO
project for which the design study is nearing completion. Several discussions took also place around the NuStorm project implementation in Europe and its physics reach and around the sterile neutrino program.
This paragraph gives a synthesis of the points which have (nearly) reached some consensus during the round table or were mentioned and so should be taken into account.

\paragraph*{General principles:}
\vspace{-10pt}
\begin{itemize}
\item Neutrino physics is very important. It is one of the fields providing evidence for "new physics".
\item It provides important information also for several theoretical grounds.
\item The link with cosmology is important and neutrino physics is used as a control piece of information.
\item The MH and CP violation are major parameters to access and should be seen as priority. An optimal program requires more than 1 experiment; need to over constrain the oscillation parameters.
\item Need 2 long baseline experiments with complementarity: Water Cerenkov detector on a few hundred km baseline, ESSnuSB or Hyper-K like) and LAr TPC on very long baseline (LBNO or LBNE like).
\item Need short/medium time scale program in neutrino, not only long term.
\item Long term project should be incremental with defined stages.
\item It is important that European nu community arrives at the generally accepted scientific program of research
\item If European contributions to neutrino physics will be realized in US- or Japan-located experiments, the needed size of the contributions will require formal international agreements.
\item Cross-section measurements should be part of the neutrino program.
\end{itemize}

\paragraph*{Sterile neutrino program:}
\vspace{-10pt}
\begin{itemize}
\item Must distinguish chasing anomalies near 1eV from searching for sterile neutrinos which can have any mass. However the anomalies at 1eV should be resolved. Confirmation of the LSND signal would completely change our picture of elementary particles.
\item It is mandatory that any new accelerator program will have synergy with LBL experiments.
\item Since there is no obvious theoretical guidance to look at them, it will be difficult to justify a project with only sterile neutrino goals.
\end{itemize}

\paragraph*{nuSTORM}
\vspace{-10pt}
\begin{itemize}
\item NuStorm is a multi-topic project which can cover steriles as well as the very important topic which is the precise measurements of cross sections, especially nue. 
\item Nustorm should be looked carefully in the CENF. The CENF designs should allow upgrading to a NuSTORM front end.
\end{itemize}

\paragraph*{Neutrino infrastructure:}
\vspace{-10pt}
\begin{itemize}
\item Neutrino program started at CERN. The steps are being defined.
\item CENF should be seen as the infrastructure, with experiments a separate consideration.
\item The community should support a program for detector R\&D and critical measurements like cross sections and hadron production. The CERN neutrino platform is an important basis to fulfil these goals. 
\item A design study about the possibility of adding to ESS the capability of generating neutrino beams is supported by the community
\end{itemize}

\end{document}